\documentclass[12pt]{article}
\usepackage{amssymb,amsmath,epsfig}
\allowdisplaybreaks

\begin{document}
\title{\bf Cosmological Consequences of New Dark Energy Models in Einstein-Aether Gravity}
\author{Shamaila Rani$^1$ \thanks{shamailatoor.math@yahoo.com}, Abdul
Jawad $^1$
\thanks{jawadab181@yahoo.com; abduljawad@cuilahore.edu.pk},
Kazuharu Bamba $^2$ \thanks{bamba@sss.fukushima-u.ac.jp}\\ and Irfan
Ullah Malik $^1$
\thanks{malik\_irfan22@yahoo.com}\\
$^1$ Department of Mathematics, COMSATS University\\ Islamabad,
Lahore-Campus, Lahore-54000, Pakistan.\\
$^2$ Division of Human Support System, Faculty of Symbiotic\\
Systems Science, Fukushima University, Fukushima 960-1296,\\ Japan.}

\date{}
\maketitle
\begin{abstract}
In this paper, we reconstruct various solutions for the accelerated
universe in the Einstein-Aether theory of gravity. For this purpose,
we obtain the effective density and pressure for Einstein-Aether
theory. We reconstruct the Einstein-Aether models by comparing its
energy density with various newly proposed holographic dark energy
models such as Tsallis, R$\acute{e}$nyi and Sharma-Mittal. For this
reconstruction, we use two forms of scale factor, power-law and
exponential forms. The cosmological analysis of underlying scenario
has been done by exploring different cosmological parameters. This
includes equation of state parameter, squared speed of sound and
evolutionary equation of state parameter via graphical
representation. We obtain some favorable results for some values of
model parameters.
\end{abstract}

\section{Introduction}

Nowadays, it is believed that our universe undergoes an accelerated
expansion with the passage of cosmic time. This cosmic expansion has
confirmed through various observational schemes such as Supernova
type Ia (SNIa) \cite{1}-\cite{4} and Cosmic Microwave Background
(CMB) \cite{5}-\cite{9}. The source behind the expansion of the
universe is a mysterious force called dark energy (DE) and its
nature is still ambiguous \cite{10}-\cite{13}. The current Planck
data shows that there is $68.3\%$ DE of the total energy contents of
the universe. The first candidate for describing DE phenomenon is
cosmological constant but it has fine tuning and cosmic coincidence
problems. Due to this reason, different DE models as well as
theories of gravity with modifications have been suggested. The
dynamical DE models include a family of Chaplygin gas as well as
holographic DE models, scalar field models such as K-essence,
phantom, quintessence, ghost etc \cite{17,17a,17aa}.


One of the DE model is holographic DE (HDE) model which becomes a
favorable technique now-a-days to study the DE mystery. This model
is established in the framework of holographic principle which
corresponds to the area instead of volume for the scaling of number
of degrees of freedom of a system. This model is an interesting
effort in exploring the nature of DE in the framework of quantum
gravity. In addition, HDE model gives the relationship between the
energy density of quantum fields in vacuum (as the DE candidate) to
the cutoffs (infrared and ultraviolet). Cohen et al. \cite{33ss}
provided a very useful result about the expression of HDE model
density which is based on the vacuum energy of the system. The black
hole mass should not overcome by the maximum amount of the vacuum
energy. Taking into account the nature of spacetime along with long
term gravity, various entropy formalism have been used to discuss
the gravitational and cosmological setups
\cite{29ss,28ss,30ss,24ss}. Recently, some new HDE models are
proposed like Tsallis HDE (THDE) \cite{28ss}, R\'{e}nyi HDE model
(RHDE) \cite{30ss} and Sharma-Mittal HDE (SMHDE) \cite{24ss}.

The examples of theories with modification setups include $f(R),
f(T)$, $f(R,\mathcal{T}),~f(G)$ etc where $R$ shows the Ricci scalar
representing the curvature, $T$ means the torsion scalar,
$\mathcal{T}$ is the trace of the energy-momentum tensor and $G$
goes as the invariant of Gauss-Bonnet \cite{22}-\cite{31}. For
recent reviews in terms of DE problem including modified gravity
theories, see, for instance~\cite{Nojiri:2010wj, Capozziello:2011et,
Capozziello:2010zz, Bamba:2015uma, Cai:2015emx, Nojiri:2017ncd,
Bamba:2012cp}. Einstein-Aether theory is one of the modified theory
of gravity \cite{32,33} and accelerated expansion phenomenon of the
universe has also been investigated in this theory \cite{34}. Meng
et al. have also discussed the current cosmic acceleration through
DE models in this gravity \cite{44,45}. Recently, Pasqua et al.
\cite{46} have made versatile study on cosmic acceleration through
various cosmological models in the presence of HDE models.


In the present work, we will develop the Einstein-Aether gravity
models in the presence of modified HDE models and well-known scale
factors. For these models of modified gravity, we will extract
various cosmological parameters. In the next section, we will brief
review of Einstein-Aether theory. In section \textbf{3}, we present
the basic cosmological parameters as well as well-known scale
factors. We will discuss the cosmological parameters for modfied HDE
models in sections \textbf{4, 5} and \textbf{6}. In the last, we
will summarize our results.

\section{Einstein-Aether Theory}

As our universe is full with many of the natural occurring
phenomenons. One of them is transfer of light from one place to
another and second is how gravity acts. To explain these kinds of
phenomenons, many of the physicists were used the concept of Aether
in many of the theories. In modern physics, Aether indicates a
physical medium that is spread homogeneously at each point of the
universe. Hence, it was considered that it is a medium in space that
helps light to travel in a vacuum. According to this concept, a
particular static frame reference is provided by Aether and
everything has absolute relative velocity in this frame. That is
suitable for Newtonian dynamics extremely well. But, when Einstein
performed different experiments on optics in his theory of
relativity, then Einstein rejected this ambiguity.
When CMB was
introduced, many of the people took it a modern form of Aether.
Gasperini has popularized Einstein-Aether theories \cite{47}. This
theory is said to be covariant modification of general relativity in
which unit time like vector field(aether) breaks the Lorentz
Invariance (LI) to examine the gravitational and cosmological
effects of dynamical preferred frame \cite{48}. Following is the
action of Einstein-Aether theory \cite{49,50}.
\begin{equation}\label{1}
S=\int d^{4}x\sqrt{-g}\bigg(\frac{R}{4\pi G} + L_{EA} + L_m\bigg),
\end{equation}
where $L_{EA}$ represents the Lagrangian density for the vector
field and $L_m$ indicates Lagrangian density of matter field.
Further, $g,~R$ and $G$ indicate determinant of the metric tensor
$g^{\mu\nu}$, Ricci scalar and gravitational constant respectively.
The Lagrangian density for vector field can be written as
\begin{eqnarray}\label{2}
L_{EA}&=&\frac{M^2}{16\pi G}F(K) + \frac{1}{16\pi G}\lambda (A^a A_a
+ 1), \\\label{3} K&=&
M^{-2}K^{ab}_{cd}\nabla_{a}A^{c}\nabla_{b}A^{d} c\\\label{4}
K^{ab}_{cd}&=& c_{1}g^{ab}g_{cd} + c_{2}\delta^{a}_{c}\delta^{b}_{d}
+ c_{3}\delta^{a}_{d}\delta^{b}_{c},\quad a, b = 0, 1, 2, 3.
\end{eqnarray}
where $\lambda$ represents a Lagrangian multiplier, dimensionless
constants are denoted by $c_i,~M$ referred as coupling constant
parameter and $A^a$ is a tensor of rank one, that is a vector. The
function $F(K)$ is any arbitrary function of $K$. We obtain the
Einstein field equations from Eq.(\ref{1}) for Einstein-Aether
theory as follows
\begin{eqnarray}\label{5}
G_{ab}&=&T^{EA}_{ab} + 8\pi G T^{m}_{ab}, \\\label{6}
\nabla_{a}\bigg(\frac{dF}{dK}J^{a}_{b}\bigg)&=& 2\lambda A_{b},
\end{eqnarray}
where $J^{a}_{b}=-2K^{ad}_{bc}\nabla_{d}A^{c},~T^{EA}_{ab}$ shows
energy momentum-tensor for vector field and $T^{m}_{ab}$ indicates
energy-momentum tensor for mater field. These tensors are given as
\begin{eqnarray}\label{9}
T^{m}_{ab}&=&(\rho + p)u_{a}u{b} + pg_{ab}, \\\nonumber
T^{EA}_{ab}&=&\frac{1}{2}\nabla_{d}\bigg((J^{d}_{a}A_{b}-J^{d}_{a}A_{b}-J_{(ab)}A^{d})\frac{dF}{dK}\bigg)-Y_{(ab)}\frac{dF}{dK}
\\\label{10}&+& \frac{1}{2}g_{ab}M^{2}F + \lambda A_{a}A_{b},
\end{eqnarray}
where $p$ and $\rho$ represent energy density and pressure of the
matter respectively. Furthermore, $u_{a}$ expresses the
four-velocity vector of the fluid and given as $u_{a}=(1, 0, 0, 0)$
and $A_{a}$ is time-like unitary vector and is defined as $A_{a}=(1,
0, 0, 0)$. Moreover $Y_{ab}$ is defined as
\begin{equation}\label{11}
Y_{ab}=
c_{1}\bigg((\nabla_{d}A_{a})(\nabla^{d}A_{b})-(\nabla_{a}A_{d})(\nabla_{a}A^{d})\bigg),
\end{equation}
where indices $(a~~b)$ show the symmetry.

The Friedmann equations modified by the Einstein-Aether gravity are
given as follows
\begin{eqnarray}\label{13}
\epsilon\bigg(\frac{F}{2K}-\frac{dF}{dK}\bigg)H^{2}+\bigg(H^{2}+\frac{k}{a^2}\bigg)&=&\bigg(\frac{8\pi
G}{3}\bigg)\rho, \\\label{14}
\epsilon\frac{d}{dt}\bigg(H\frac{dF}{dK}\bigg)+\bigg(-2\dot{H}+\frac{2k}{a^2}\bigg)&=&8\pi
G(p+\rho).
\end{eqnarray}
Here $K$ becomes $K=\frac{3\epsilon H^2}{M^2}$, where $\epsilon$ is
a constant parameter. The energy density of Einstein-Aether theory
is denoted by $\rho_{EA}$ and called effective energy density while
the effective pressure in Einstein-Aether gravity is given by
$p_{EA}$. So, we can rewrite Eqs.(\ref{13}) and (\ref{14}) as
\begin{eqnarray}\label{15}
\bigg(H^{2}+\frac{k}{a^2}\bigg)&=&\bigg(\frac{8\pi G}{3}\bigg)\rho +
\frac{1}{3} \rho_{EA}, \\\label{16}
\bigg(-2\dot{H}+\frac{2k}{a^2}\bigg)&=&8\pi
G(p+\rho)+(\rho_{EA}+p_{EA}),
\end{eqnarray}
where
\begin{eqnarray}\label{17}
\rho_{EA}&=&3\epsilon
H^{2}\bigg(\frac{dF}{dK}-\frac{F}{2K}\bigg),\\\label{18}
p_{EA}&=&-3\epsilon
H^{2}\bigg(\frac{dF}{dK}-\frac{F}{2K}\bigg)-\epsilon\bigg(\dot{H}\frac{dF}{dK}+H\frac{d\dot{F}}{dK}\bigg).
\\\label{19}&=&\rho_{EA}-\frac{\dot{\rho}_{EA}}{3H}.
\end{eqnarray}
The EoS parameter for Einstein-Aether can be obtained by using
Eqs.(\ref{17}) and (\ref{18}), which is given by
\begin{equation}\label{21}
\omega_{EA}=\frac{p_{EA}}{\rho_{EA}}=-1-\frac{\dot{H}\frac{dF}{dK}+H\frac{d\dot{F}}{dK}}{3H^{2}(\frac{dF}{dK}-\frac{F}{2K})}.
\end{equation}

\section{Cosmological Parameters}

To understand the geometry of the universe, following are some basic
cosmological parameters.

\subsection{Equation of State Parameter}

In order to categorize the different phases of the evolving
universe, the EoS parameter is widely used. In particular, the
decelerated and accelerated phases which contain DE, DM, radiation
dominated eras. This parameter is defined in terms of energy density
$\rho$ and pressure $p$ as $\omega=\frac{p}{\rho}$.
\begin{itemize}
\item In the decelerated phase, the radiation era
$0<\omega<\frac{1}{3}$ and cold DM era $\omega=0$ are included.
\item The accelerated phase of the universe has following eras: $\omega=-1~\Rightarrow$ cosmological
constant, $-1<\omega<\frac{-1}{3}~\Rightarrow$ quintessence and
 $\omega<-1~\Rightarrow$ phantom era of the universe.
\end{itemize}

\subsection{Squared Speed of Sound}

To examine the behavior of DE models, there is another parameter
which is known as squared speed of sound. It is denoted by $v_s^2$
and is calculated by the following formula
\begin{equation}
v_s^2=\frac{\dot{p}}{\dot{\rho}}.
\end{equation}
The stability of the model can be checked by it. If its graph is
showing negative values then we may say that model is unstable and
in case of non-negative values of the graph, it represents the
stable behavior of the model.

\subsection{$\omega$-$\omega^\prime$ Plane}

There are different DE models which have different properties. To
examine their dynamical behavior, we use $\omega$-${\omega}'$ plane,
where prime denotes the derivative with respect to $\ln a$ and
subscript $\Lambda$ indicates DE scenario. This method was developed
by Caldwell and Linder \cite{51} and dividing $\omega$-${\omega}'$
plane into two parts. One is the freezing part in which evolutionary
parameter gives negative behavior for negative EoS parameter, i.e.,
(${\omega}'<0, \omega<0$) while for positive behavior of
evolutionary parameter corresponding to negative EoS parameter
yields thawing part (${\omega}' > 0, \omega < 0$) of the evolving
universe.

\subsection{Scale Factor}

The scale factor is the measure that how much the universe has
expanded since given time. It is represented by $a(t)$. Since, the
latest cosmic observations have shown that universe is accelerating
so $a(t)>0$. As Einstein-Aether is one of the modified theory which
may produce the accelerated expansion of the universe. By using this
theory, we can reconstruct various well-known DE models. In order to
do this, we take some modified HDE models such as THDE, RHDE and
SMHDE models. Since $F(K)$ is a function that the Einstein-Aether
theory contains, which can be determined by comparing the densities
with the above DE models. For this purpose, we use some well-known
forms of scale factor $a(t)$. We consider two forms of scale factors
$a(t)$ in terms of power and exponential terms. These
are\\
(i) \underline{\textbf{Power-law form:}} $a(t) = a_{0}t^m,~m > 0$,
where $a_0$ is a constant which indicates the value of scale factor
at present-day \cite{52}. From this scale factor, we get $H,~
\dot{H},~K$ as follows
\begin{equation}\label{23}
H = \frac{m}{t},\quad  \dot{H} = -\frac{m}{t^2},\quad   K =
\frac{3\epsilon m^2}{M^{2}t^2}.
\end{equation}
(ii) \underline{\textbf{Exponential form:}} $a(t) = e^{\alpha
t^{\theta}}$ where $\alpha$ is a positive constant and $\theta$ lies
between $0$ and $1$. This scale factor gives
\begin{equation}\label{24}
H = \alpha\theta t^{\theta-1} ,\quad \dot{H} =
\alpha\theta(\theta-1)t^{\theta-2} ,\quad K = \frac{3\epsilon
\alpha^{2}\theta^{2}t^{2(\theta-1)}}{M^2}.
\end{equation}

\section{Reconstruction from Tsallis Holographic Dark Energy Model}

The energy density of THDE model is given by \cite{28ss}
\begin{equation}\label{26}
\rho_D=BL^{2\delta-4},
\end{equation}
where $B$ is an unknown parameter. Taking into account Hubble radius
as IR cutoff $L$, that is $L=\frac{1}{H}$, we have
\begin{equation}\label{27}
\rho_D=BH^{-2\delta+4}.
\end{equation}
In order to construct a DE model in the framework of Einstein-Aether
gravity with THDE model, we compare the densities of both models,
(i.e., $\rho_{EA}=\rho_D$). This yields
\begin{equation}\label{28}
\frac{dF}{dK}-\frac{F}{2K}=\frac{B}{3\epsilon}H^{-2\delta+2},
\end{equation}
which results the following form
\begin{equation}\label{29}
F(K)=\frac{2BM^{-2\delta+2}K^{-\delta+2}}{(-2\delta+3)(3\epsilon)^{-\delta+2}}+C_1\sqrt{K}.
\end{equation}
\underline{\textbf{Power-law form of scale factor:}}\\ Using the
expression of $F(K)$ along with Eq.(\ref{23}) in (\ref{17}), we
obtain the energy density and pressure as
\begin{eqnarray}\label{30}
\rho_{EA}&=&\frac{m^2 \bigg(3^{\delta }2 B K^{\frac{5}{2}-\delta }
M^{2-2 \delta } \epsilon ^{\delta }-9 \epsilon ^2 C_1\bigg)}{6
K^\frac{3}{2} t^2 \epsilon }, \\\nonumber p_{EA}&=&m \bigg(4 B M^{-2
\delta } \epsilon ^{\delta } \bigg(-3^{\delta } K^{1-\delta } M^2
(4-9 m-2 \delta +6 m \delta )-12 M^2 (-2\\\nonumber&+&\delta )
(-1+\delta ) \bigg(\frac{m^2 \epsilon }{M^2 t^2}\bigg)^{1-\delta
}\bigg)(-3+2 \delta)^{-1}+\bigg(\frac{9 (-1+6 m) \epsilon
^2}{K^\frac{3}{2}}+\frac{3 \sqrt{3}}{m^4}\\\nonumber&\times& M^4 t^4
\sqrt{\frac{m^2 \epsilon }{M^2 t^2}}\bigg) C_1\bigg)(36 t^2 \epsilon
)^{-1}.
\end{eqnarray}
Using these expressions of energy density and pressure, we find the
values of some cosmological parameters in the following. The EoS
parameter takes the following form
 \begin{eqnarray}\nonumber
\omega_{EA}&=&K^\frac{3}{2} \bigg(4 B M^{-2 \delta } \epsilon
^{\delta } \bigg(-3^{\delta } K^{1-\delta } M^2 (4-9 m-2 \delta +6 m
\delta )-12 M^2 (-2\\\nonumber&+&\delta ) (-1+\delta )
\bigg(\frac{m^2 \epsilon }{M^2 t^2}\bigg)^{1-\delta }\bigg)(-3+2
\delta)^1+\bigg(\frac{9 (-1+6 m) \epsilon ^2}{K^\frac{3}{2}}+\frac{3
\sqrt{3} M^4 }{m^4}\\\label{100}&\times&t^4 \sqrt{\frac{m^2 \epsilon
}{M^2 t^2}}\bigg) C_1\bigg)\bigg(6 m \bigg(2 3^{\delta } B
K^{\frac{5}{2}-\delta } M^{2-2 \delta } \epsilon ^{\delta }-9
\epsilon ^2 C_1\bigg)\bigg)^{-1}.
\end{eqnarray}
The derivative of EoS parameter with respect to $\ln a$ is given by
\begin{eqnarray}\nonumber
{\omega'_{EA}}&=&K^\frac{3}{2} t \bigg(\frac{96 B m^2 M^{-2 \delta }
(1-\delta ) (-2+\delta ) (-1+\delta ) \epsilon ^{1+\delta }
\bigg(\frac{m^2 \epsilon }{M^2 t^2}\bigg)^{-\delta }}{t^3 (-3+2
\delta )}+\bigg(-\frac{3 \sqrt{3}}{m^2}\\\nonumber&\times&\frac{ M^2
t \epsilon }{ \sqrt{\frac{m^2 \epsilon }{M^2 t^2}}}+\frac{12
\sqrt{3} M^4 t^3 \sqrt{\frac{m^2 \epsilon }{M^2 t^2}}}{m^4}\bigg)
C_1\bigg)\bigg(6 m^2 \bigg(2 3^{\delta } B K^{\frac{5}{2}-\delta }
M^{2-2 \delta } \epsilon ^{\delta }-9 \epsilon ^2
C_1\bigg)\bigg)^{-1}.\\\label{101}
\end{eqnarray}
We plot EoS parameter versus $z$ using the relation
$t=\frac{1}{(1+z)^{\frac{1}{m}}}$ taking values of constants as
$B=5,~M=5,~\delta=1.8,~\epsilon=1$ and $C_1=2$. We plot
$\omega_{EA}$ for three different values of scale factor parameter
$m$ as $m=2,~3,4$ as shown in Figure 1. All the three trajectories
represent the phantom behavior of the universe related to redshift
parameter. In Figure 2, we plot $\omega'_{EA}-\omega{EA}$ plane
taking same values of the parameters for $-1\leq z \leq 1$. For
$m=2$ and 3, the evolving EoS parameter shows negative behavior with
respect to negative EoS parameter which indicates the freezing
region of the universe. The trajectory of $\omega'_{EA}$ for $m=4$
represents the positive behavior for negative EoS parameter and
expresses the evolving universe in thawing region of the universe.
\begin{figure}[h]
\begin{minipage}{14pc}
\includegraphics[width=16pc]{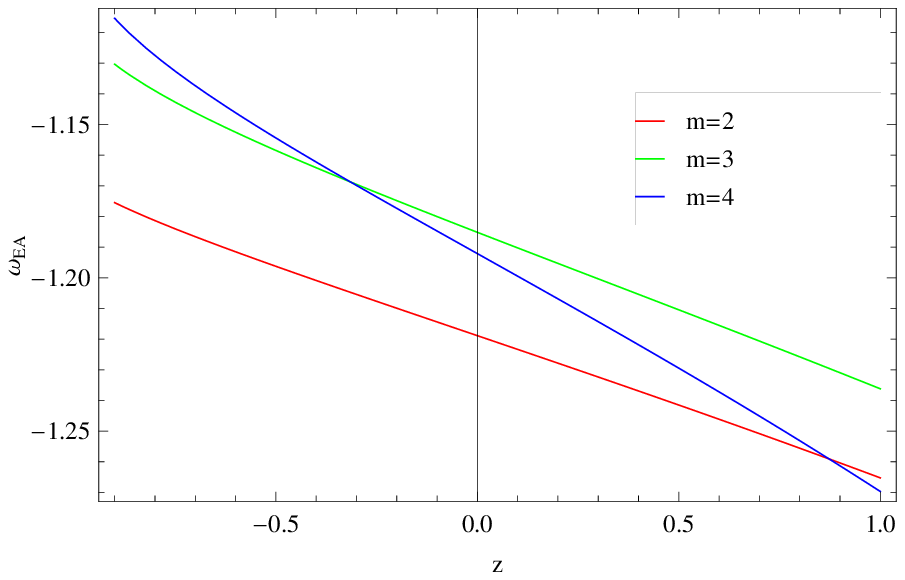}\caption{Plot of
$\omega_{EA}$ versus $z$ taking power-law scale factor for THDE
model.}
\end{minipage}\hspace{3pc}
\begin{minipage}{14pc}
\includegraphics[width=16pc]{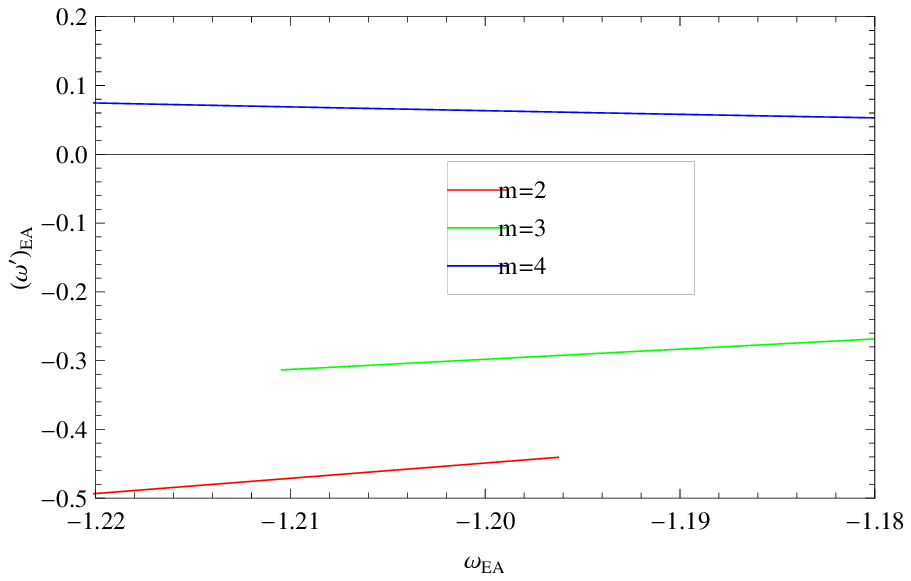}\caption{Plot of
$\omega'_{EA}-\omega_{EA}$ taking power-law scale factor for THDE
model.}
\end{minipage}\hspace{3pc}
\begin{minipage}{14pc}
\includegraphics[width=16pc]{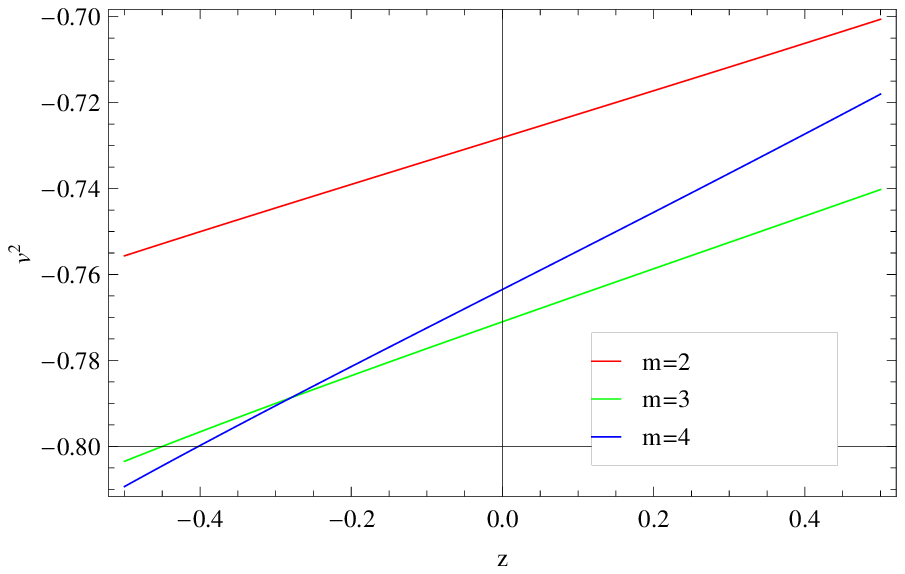}\caption{Plot of $v_s^2$
versus $z$ taking power-law scale factor for THDE model.}
\end{minipage}\hspace{3pc}
\end{figure}

Also the squared speed of sound in the underlying scenario becomes
\begin{eqnarray}\nonumber
v_s^2&=&\bigg(\bigg(\frac{m^2 \epsilon }{M^2 t^2}\bigg)^{-\delta }
\bigg(8 B m^4 \epsilon ^{\delta } \bigg(-12 K^{\frac{3}{2}+\delta }
m^2 (-2+\delta )^2 (-1+\delta ) \epsilon +3^{\delta } K^\frac{5}{2}
M^2\\\nonumber&\times&t^2 (4-9 m-2 \delta +6 m \delta )
\bigg(\frac{m^2 \epsilon }{M^2 t^2}\bigg)^{\delta }\bigg)+3
K^{\delta } M^{2 \delta } t^2 (-3+2 \delta ) \bigg(\frac{m^2
\epsilon }{M^2 t^2}\bigg)^{\delta }\\\nonumber&\times&\bigg(6 (1-6
m) m^4 \epsilon ^2+\sqrt{3} K^\frac{3}{2} M^4 t^4 \sqrt{\frac{m^2
\epsilon }{M^2 t^2}}\bigg) C_1\bigg)\bigg)/\bigg(12 m^5 t^2 (-3+2
\delta )\\\label{102}&\times&\bigg(-2 3^{\delta } B K^\frac{5}{2}
M^2 \epsilon ^{\delta }+9 K^{\delta } M^{2 \delta } \epsilon ^2
C_1\bigg)\bigg).
\end{eqnarray}
Figure 3 shows the plot of $v^2_s$ versus $z$ to check the behavior
of Einstein-Aether model for THDE and power-law scale factor for
same values of parameters. The trajectories represent the negative
behavior of the model which indicated the instability of
the model.\\
\underline{\textbf{Exponential form of scale factor:}}\\
Following the same steps for exponential form of scale factor, we
get energy density and pressure as
\begin{eqnarray}\label{36}
\rho_{EA}&=&\frac{t^{-2+2 \theta } \alpha ^2 \theta ^2 \bigg(2
3^{\delta } B K^{\frac{5}{2}-\delta } M^{2-2 \delta } \epsilon
^{\delta }-9 \epsilon ^2 C_1\bigg)}{6 K^\frac{3}{2} \epsilon },
\\\nonumber p_{EA}&=&\frac{1}{12 \epsilon }t^{-2+\theta } \alpha
\theta \bigg(\frac{2 t^{\theta } \alpha  \theta  \bigg(-2 3^{\delta
} B K^{\frac{5}{2}-\delta } M^{2-2 \delta } \epsilon ^{\delta }+9
\epsilon ^2 C_1\bigg)}{K^\frac{3}{2}}-\epsilon ^2
(-1\\\nonumber&+&\theta ) \bigg(8 B M^{-2 \delta } (-2+\delta )
\epsilon ^{-2+\delta } \bigg(3^{\delta } K^{1-\delta } M^2-6 M^2
(-1+\delta ) \bigg(\frac{t^{-2+2 \theta }
}{M^2}\\\nonumber&\times&\alpha ^2 \epsilon \theta
^2\bigg)^{1-\delta }\bigg)\bigg(3 (-3+2 \delta
)\bigg)^{-1}+\bigg(-\frac{3}{K^\frac{3}{2}}+\frac{
\sqrt{\frac{t^{-2+2 \theta } \alpha ^2 \epsilon \theta
^2}{M^2}}}{\alpha ^4 \epsilon ^2 \theta ^4}\sqrt{3}
M^4\\\nonumber&\times&t^{4-4 \theta }\bigg) C_1\bigg)\bigg).
\end{eqnarray}
Now by using above density and pressure, we obtain the EoS parameter
and its derivative for Einstein-Aether gravity as follows
\begin{eqnarray}\nonumber
\omega_{EA}&=&\bigg(K^\frac{3}{2} t^{-\theta } \bigg(\frac{2
t^{\theta } \alpha \theta  \bigg(-2 3^{\delta } B
K^{\frac{5}{2}-\delta } M^{2-2 \delta } \epsilon ^{\delta }+9
\epsilon ^2 C_1\bigg)}{K^\frac{3}{2}}-\epsilon ^2 (-1+\theta )
\\\nonumber&\times&\bigg(8 B M^{-2 \delta } (-2+\delta )
\epsilon ^{-2+\delta } \bigg(3^{\delta } K^{1-\delta } M^2-6 M^2
(-1+\delta ) \bigg(\frac{t^{-2+2 \theta } \alpha ^2
}{M^2}\\\nonumber&\times&\epsilon \theta ^2\bigg)^{1-\delta
}\bigg)\bigg(3 (-3+2 \delta
)\bigg)^{-1}+\bigg(-\frac{3}{K^\frac{3}{2}}+\frac{\sqrt{3} M^4
t^{4-4 \theta } \sqrt{\frac{t^{-2+2 \theta } \alpha ^2 \epsilon
\theta ^2}{M^2}}}{\alpha ^4 \epsilon ^2 \theta
^4}\bigg)\\\nonumber&\times& C_1\bigg)\bigg)\bigg)/\bigg(2 \alpha
\theta  \bigg(2 3^{\delta } B K^{\frac{5}{2}-\delta } M^{2-2 \delta
} \epsilon ^{\delta }-9 \epsilon ^2 C_1\bigg)\bigg), \\\nonumber
{\omega'_{EA}}&=&\frac{1}{6 \alpha ^2 \theta  \bigg(2 3^{\delta } B
K^{\frac{5}{2}-\delta } M^{2-2 \delta } \epsilon ^{\delta }-9
\epsilon ^2 C_1\bigg)}K^\frac{3}{2} t^{-2 (1+\theta )} (-1+\theta )
\bigg(8 B K^{-\delta }\\\nonumber&\times&M^{-2 \delta } (-2+\delta )
\epsilon ^{\delta } \bigg(\frac{t^{-2+2 \theta } \alpha ^2 \epsilon
\theta ^2}{M^2}\bigg)^{-\delta } \bigg(-6 K^{\delta } t^{2 \theta }
\alpha ^2 (-1+\delta ) \epsilon  (2+2 \delta\\\nonumber&\times&
(-1+\theta )-\theta ) \theta +3^{\delta } K M^2 t^2
\bigg(\frac{t^{-2+2 \theta } \alpha ^2 \epsilon  \theta
^2}{M^2}\bigg)^{\delta }\bigg)\bigg(-3+2 \delta
\bigg)^{-1}-3\\\nonumber&\times&t^{2-4 \theta } \bigg(3 t^{4 \theta
} \alpha ^4 \epsilon ^2 \theta ^5+\sqrt{3} K^\frac{3}{2} M^4 t^4
(3-4 \theta ) \sqrt{\frac{t^{-2+2 \theta } \alpha ^2 \epsilon \theta
^2}{M^2}}\bigg) C_1\bigg(K^\frac{3}{2} \alpha ^4\\\nonumber&\times&
\theta ^5\bigg)^{-1}\bigg).
\end{eqnarray}
\begin{figure}[h]
\begin{minipage}{14pc}
\includegraphics[width=16pc]{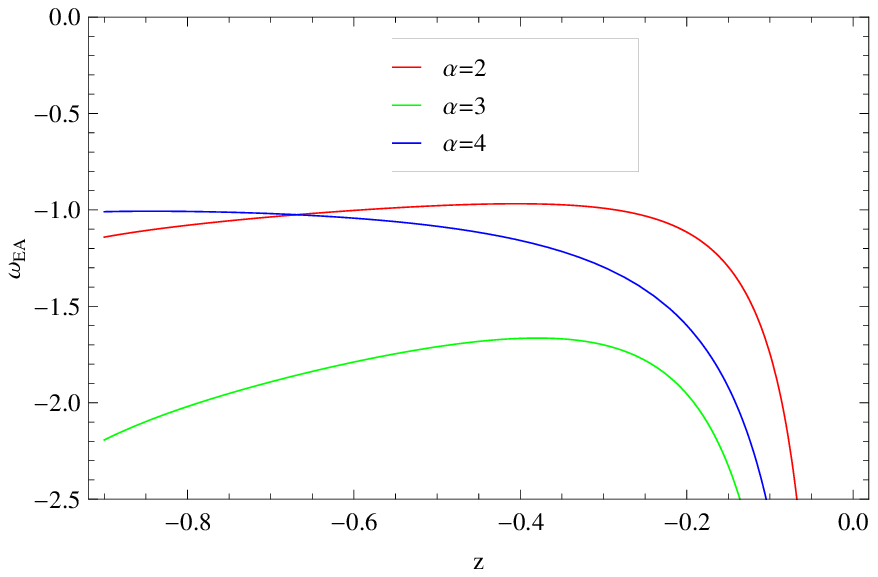}\caption{Plot of
$\omega_{EA}$ versus $z$ taking exponential scale factor for THDE
model.}
\end{minipage}\hspace{3pc}
\begin{minipage}{14pc}
\includegraphics[width=16pc]{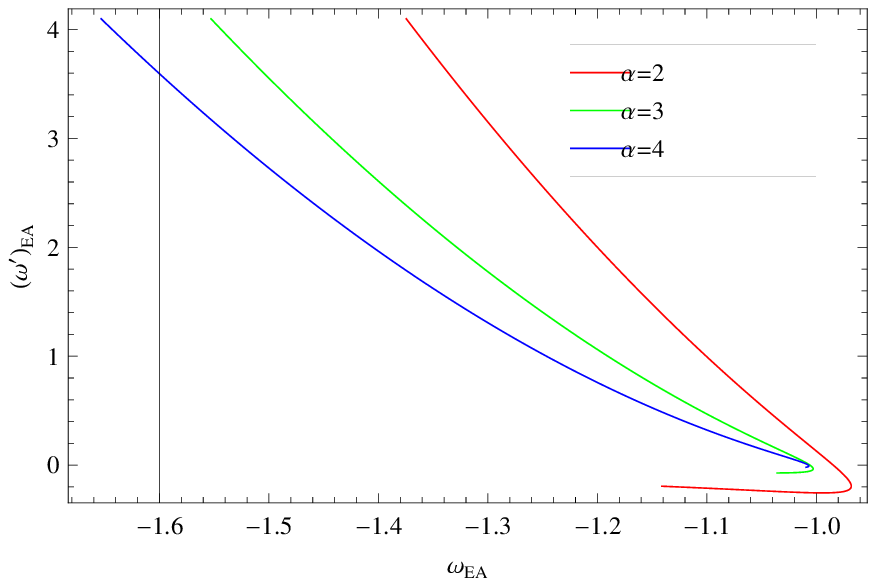}\caption{Plot of
$\omega'_{EA}-\omega_{EA}$ taking exponential scale factor for THDE
model.}
\end{minipage}\hspace{3pc}
\begin{minipage}{14pc}
\includegraphics[width=16pc]{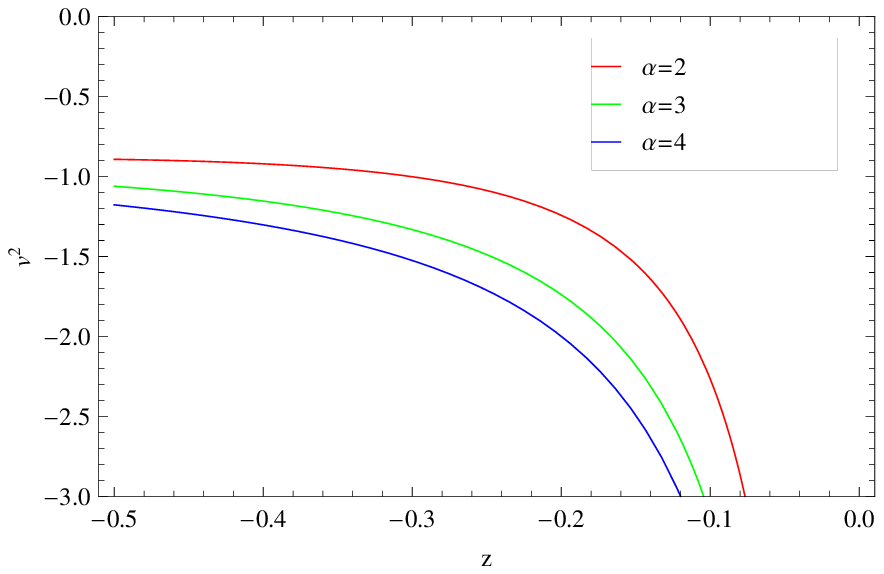}\caption{Plot of $v_s^2$
versus $z$ taking exponential scale factor for THDE model.}
\end{minipage}\hspace{3pc}
\end{figure}

The squared speed of sound for second form of scale factor is given
by
\begin{eqnarray}\nonumber
v_s^2&=&K^\frac{3}{2} t^{-\theta } \bigg(8 B M^{-2 \delta } \epsilon
^{\delta } \bigg(-6 M^2 (-2+\delta ) (-1+\delta ) (4+2 \delta
(-1+\theta )-3 \theta )\\\nonumber&\times&\bigg(\frac{t^{-2+2 \theta
} \alpha ^2 \epsilon  \theta ^2}{M^2}\bigg)^{1-\delta }-3^{\delta }
K^{1-\delta } M^2 \bigg((-2+\delta ) (-2+\theta )+3 t^{\theta }
\alpha  (-3+2\\\nonumber&\times& \delta ) \theta \bigg)\bigg)(-3+2
\delta )^{-1}+3 t^{-4 \theta } \bigg(3 t^{4 \theta } \alpha ^4
\epsilon ^2 (-2+\theta ) \theta ^4+36 t^{5 \theta } \alpha ^5
\epsilon ^2 \theta ^5+\sqrt{3}\\\nonumber&\times& K^\frac{3}{2} M^4
t^4 \sqrt{\frac{t^{-2+2 \theta } \alpha ^2 \epsilon  \theta
^2}{M^2}} (-1+2 \theta )\bigg) C_1(K^\frac{3}{2} \alpha ^4 \theta
^4)^{-1}\bigg)\bigg(12 \alpha  \theta \bigg(2 3^{\delta }
B\\\nonumber&\times& K^{\frac{5}{2}-\delta } M^{2-2 \delta }
\epsilon ^{\delta }-9 \epsilon ^2 C_1\bigg)\bigg).
\end{eqnarray}

Figure 4 represents the graph of EoS parameter versus $z$ for
exponential form of scale factor taking
$B=5=M,~\delta=1.8,~\epsilon=1,C_1=-0.5,~\theta=0.5$ and scale
factor parameter $\alpha=2,~3,~4$. This parameter represents the
phantom behavior of the universe for $\alpha=3$ and after a
transition from quintessence to phantom era for $\alpha=2$. For
$\alpha=4$, the trajectory of the EoS parameter corresponds to the
$\Lambda$CDM model $\omega_{EA}=-1$. In Figure 5, the graph is
plotted between $\omega'_{EA}$ and $\omega_{EA}$. The graph
represents initially freezing region and then indicates the thawing
region of the evolving universe. As we increase the value of
$\alpha$, the trajectories indicate the thawing region only.
However, the graph of $v^2_s$ versus $z$ as shown in Figure 6 shows
the unstable behavior.

\section{Reconstruction from R$\acute{e}$nyi Holographic Dark Energy Model}

The energy density of  RHDE model is \cite{24ss}
\begin{equation}\label{39}
\rho_D=\frac{3C^{2}L^{-2}}{8\pi\bigg(1+\frac{\delta\pi}{H^2}\bigg)}.
\end{equation}
For Hubble horizon, it takes the form
\begin{equation}\label{40}
\rho_D=\frac{3C^{2}H^{2}}{8\pi\bigg(1+\frac{\delta\pi}{H^2}\bigg)}.
\end{equation}
Now we compare the Einstein-Aether model energy density with the
RHDE model density (i.e., $\rho_{EA}=\rho_D$) in order to get
reconstructed equation,
\begin{equation}\label{41}
\frac{dF}{dK}-\frac{F}{2K}=\frac{C^2}{8\pi\epsilon(1+\frac{\delta\pi}{H^2})}.
\end{equation}
The solution of this equation is given by
\begin{equation}\label{42}
F(K)=\frac{C^2}{4\pi M\epsilon}\bigg(KM-\sqrt{3K\delta \epsilon \pi
}\arctan\bigg(\frac{M\sqrt{K}}{\sqrt{3\epsilon \pi
\delta}}\bigg)\bigg)+C_2\sqrt{K}.
\end{equation}
\underline{\textbf{Power-law form of scale factor:}}\\
Inserting all the corresponding values in Eqs.(\ref{17}) and
(\ref{18}), we get density and pressure of Einstein-Aether gravity
model as follows
\begin{eqnarray}\nonumber
\rho_{EA}&=&\frac{1}{8 \pi  t^2}3 C^2 m^2 \bigg(1-\frac{3
\sqrt{\delta \epsilon \pi } \sqrt{K \delta  \epsilon  \pi
}}{\sqrt{K} \bigg(K M^2+3 \delta \epsilon  \pi \bigg)}\bigg).
\\\nonumber p_{EA}&=&\frac{1}{24 \pi t^5}C^2 m \bigg(9 m t^3
\bigg(-1+\frac{3 \sqrt{\delta  \epsilon  \pi } \sqrt{K \delta
\epsilon  \pi }}{\sqrt{K} \bigg(K M^2+3 \delta \epsilon  \pi
\bigg)}\bigg)-\frac{1}{M^4}\epsilon  \bigg(3 M^3\\\nonumber&\times&
t^3 \bigg(M \bigg(-2+\frac{3 \sqrt{\delta \epsilon  \pi } \sqrt{K
\delta \epsilon  \pi }}{\sqrt{K} \bigg(K M^2+3 \delta  \epsilon  \pi
\bigg)}\bigg)+\frac{\text{ArcTan}\bigg(\frac{\sqrt{K} M}{\sqrt{3}
\sqrt{\delta  \epsilon \pi }}\bigg)}{K}\sqrt{3}\\\nonumber&\times&
\sqrt{K \delta \epsilon  \pi }-\frac{C_2}{\sqrt{K}}\bigg)(\epsilon
)^{-1}+\bigg(m^2 \bigg(-3 m^2 t \delta ^2 \epsilon ^2 \pi ^2
\bigg(m^2-t^2 \delta \pi \bigg)-3\\\nonumber&\times& m \delta ^{3/2}
\epsilon ^2 \pi ^\frac{3}{2} \bigg(m^2+t^2 \delta \pi \bigg)^2
\text{ArcTan}\bigg(\frac{m}{t \sqrt{\delta } \sqrt{\pi
}}\bigg)+\sqrt{3} M t \sqrt{\delta  \epsilon  \pi }
m\epsilon\\\nonumber&\times&\sqrt{\frac{ \delta \pi }{M^2 t^2}}
\bigg(m^2+t^2 \delta \pi \bigg)^2
C_2\bigg)\bigg)/\bigg(\bigg(\frac{m^2 \epsilon }{M^2
t^2}\bigg)^{3/2} \sqrt{\delta  \epsilon  \pi } \sqrt{\frac{m^2
\delta  \epsilon ^2 \pi }{M^2 t^2}} \bigg(m^2\\\nonumber&+&t^2
\delta \pi \bigg)^2\bigg)\bigg)\bigg).
\end{eqnarray}
\begin{figure}[h]
\begin{minipage}{14pc}
\includegraphics[width=16pc]{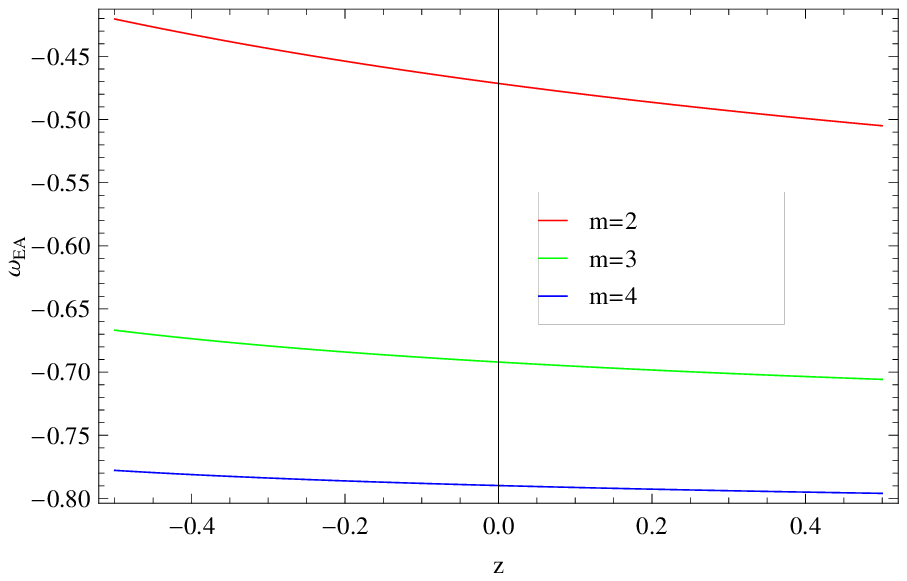}\caption{Plot of
$\omega_{EA}$ versus $z$ taking power-law scale factor for RHDE
model.}
\end{minipage}\hspace{3pc}
\begin{minipage}{14pc}
\includegraphics[width=16pc]{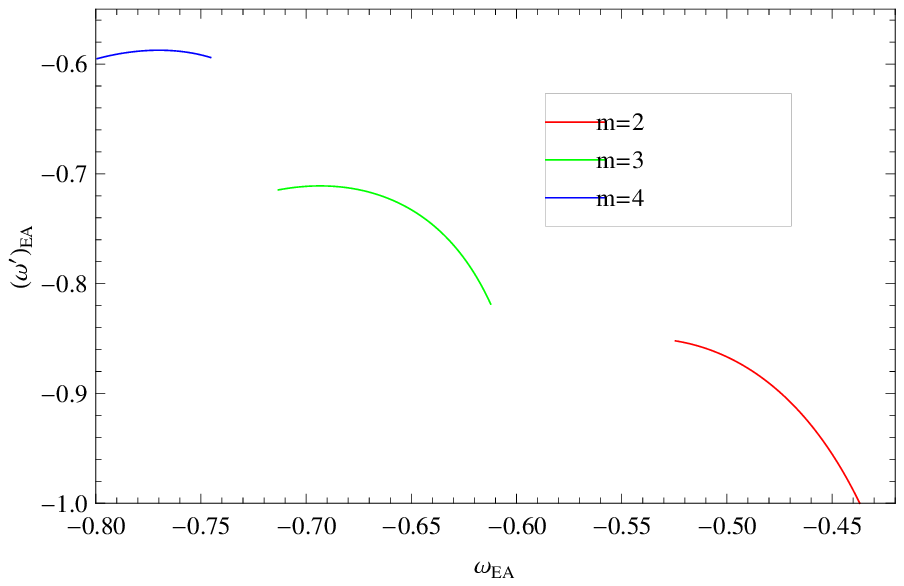}\caption{Plot of
$\omega'_{EA}-\omega_{EA}$ taking power-law scale factor for RHDE
model.}
\end{minipage}\hspace{3pc}
\begin{minipage}{14pc}
\includegraphics[width=16pc]{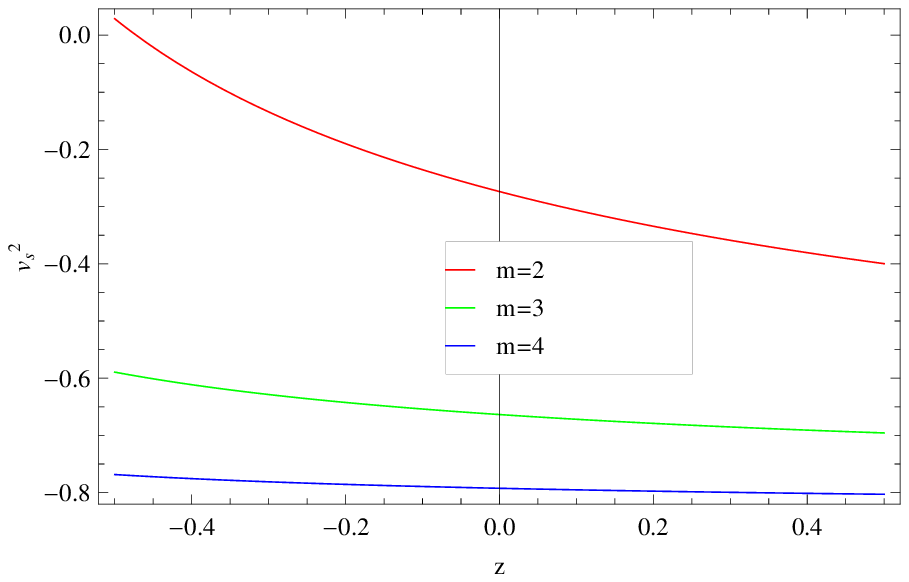}\caption{Plot of $v_s^2$
versus $z$ taking power-law scale factor for RHDE model.}
\end{minipage}\hspace{3pc}
\end{figure}
In this case, the EoS parameter takes the form
\begin{eqnarray}\nonumber
\omega_{EA}&=&\bigg(9 m t^3 \bigg(-1+\frac{3 \sqrt{\delta  \epsilon
\pi } \sqrt{K \delta  \epsilon  \pi }}{\sqrt{K} \bigg(K M^2+3 \delta
\epsilon  \pi \bigg)}\bigg)-\frac{1}{M^4}\epsilon  \bigg(3 M^3 t^3
\bigg(M \bigg(-2\\\nonumber&+&\frac{3 \sqrt{\delta  \epsilon  \pi }
\sqrt{K \delta  \epsilon  \pi }}{\sqrt{K} \bigg(K M^2+3 \delta
\epsilon  \pi \bigg)}\bigg)+\frac{\sqrt{3} \sqrt{K \delta \epsilon
\pi } \text{ArcTan}\bigg(\frac{\sqrt{K} M}{\sqrt{3} \sqrt{\delta
\epsilon \pi
}}\bigg)}{K}-\frac{C_2}{\sqrt{K}}\bigg)\\\nonumber&\times&(\epsilon
)^{-1}+\bigg(m^2 \bigg(-3 m^2 t \delta ^2 \epsilon ^2 \pi ^2
\bigg(m^2-t^2 \delta \pi \bigg)-3 m \delta ^\frac{3}{2} \epsilon ^2
\pi ^\frac{3}{2} \bigg(m^2+t^2\\\nonumber&\times& \delta \pi
\bigg)^2 \text{ArcTan}\bigg(\frac{m}{t \sqrt{\delta } \sqrt{\pi
}}\bigg)+\sqrt{3} M t \sqrt{\delta  \epsilon  \pi } \sqrt{\frac{m^2
\delta  \epsilon ^2 \pi }{M^2 t^2}} \bigg(m^2+t^2 \delta  \pi
\bigg)^2 \\\nonumber&\times&C_2\bigg)\bigg)/\bigg(\bigg(\frac{m^2
\epsilon }{M^2 t^2}\bigg)^\frac{3}{2} \sqrt{\delta  \epsilon  \pi }
\sqrt{\frac{m^2 \delta  \epsilon ^2 \pi }{M^2 t^2}} \bigg(m^2+t^2
\delta  \pi \bigg)^2\bigg)\bigg)\bigg)/\bigg(9 t^3
\bigg(m\\\nonumber&-&\frac{3 m \sqrt{\delta  \epsilon  \pi } \sqrt{K
\delta \epsilon  \pi }}{\sqrt{K} \bigg(K M^2+3 \delta  \epsilon  \pi
\bigg)}\bigg)\bigg),
\end{eqnarray}
and ${\omega'_{EA}}$ is given as follows
\begin{eqnarray}\nonumber
{\omega'_{EA}}&=&3 \bigg(6-\frac{4 \sqrt{\frac{m^2 \delta \epsilon
^2 \pi }{M^2 t^2}}}{\sqrt{\frac{m^2 \epsilon }{M^2 t^2}}
\sqrt{\delta \epsilon  \pi }}+m \bigg(-9+\frac{27 \sqrt{\delta
\epsilon  \pi } \sqrt{K \delta  \epsilon  \pi }}{\sqrt{K} \bigg(K
M^2+3 \delta \epsilon  \pi \bigg)}\bigg)+\sqrt{\delta  \epsilon  \pi
}\\\nonumber&\times& \bigg(-\frac{9 \sqrt{K \delta  \epsilon  \pi
}}{\sqrt{K} \bigg(K M^2+3 \delta  \epsilon  \pi \bigg)}+\frac{4 m^2
\sqrt{\frac{m^2 \epsilon }{M^2 t^2}} \bigg(m^4+4 m^2 t^2 \delta  \pi
+t^4 \delta ^2 \pi ^2\bigg)}{\sqrt{\frac{m^2 \delta  \epsilon ^2 \pi
}{M^2 t^2}} \bigg(m^2+t^2 \delta  \pi
\bigg)^3}\bigg)\\\nonumber&+&\frac{4 t \sqrt{\delta } \sqrt{\pi }
\sqrt{\frac{m^2 \delta  \epsilon ^2 \pi }{M^2 t^2}}
\text{ArcTan}\bigg(\frac{m}{t \sqrt{\delta } \sqrt{\pi }}\bigg)}{m
\sqrt{\frac{m^2 \epsilon }{M^2 t^2}} \sqrt{\delta \epsilon  \pi
}}-\frac{3 \sqrt{3} \sqrt{K \delta  \epsilon  \pi }
\text{ArcTan}\bigg(\frac{\sqrt{K} M}{\sqrt{3} \sqrt{\delta  \epsilon
\pi }}\bigg)}{K M}\bigg)\\\nonumber&+&\frac{\bigg(9-\frac{4 \sqrt{3}
\sqrt{K}}{\sqrt{\frac{m^2 \epsilon }{M^2 t^2}}}\bigg) C_2}{\sqrt{K}
M}\bigg(9 m \bigg(m-\frac{3 m \sqrt{\delta  \epsilon  \pi } \sqrt{K
\delta  \epsilon  \pi }}{\sqrt{K} \bigg(K M^2+3 \delta  \epsilon \pi
\bigg)}\bigg)\bigg)^{-1}.
\end{eqnarray}

The expression for squared speed of sound turns out as
\begin{eqnarray}\nonumber
v_s^2&=&-3 \bigg(-4+\frac{\sqrt{\frac{m^2 \delta  \epsilon ^2 \pi
}{M^2 t^2}}}{\sqrt{\frac{m^2 \epsilon }{M^2 t^2}} \sqrt{\delta
\epsilon  \pi }}+m \bigg(6-\frac{18 \sqrt{\delta  \epsilon  \pi }
\sqrt{K \delta  \epsilon  \pi }}{\sqrt{K} \bigg(K M^2+3 \delta
\epsilon  \pi \bigg)}\bigg)+\sqrt{\delta  \epsilon  \pi
}\\\nonumber&\times& \bigg(\frac{6 \sqrt{K \delta  \epsilon  \pi
}}{\sqrt{K} \bigg(K M^2+3 \delta  \epsilon  \pi \bigg)}-\frac{m^2
\sqrt{\frac{m^2 \epsilon }{M^2 t^2}} \bigg(m^4+4 m^2 t^2 \delta  \pi
+11 t^4 \delta ^2 \pi ^2\bigg)}{\sqrt{\frac{m^2 \delta  \epsilon ^2
\pi }{M^2 t^2}} \bigg(m^2+t^2 \delta  \pi
\bigg)^3}\bigg)\\\nonumber&-&\frac{t \sqrt{\delta } \sqrt{\pi }
\sqrt{\frac{m^2 \delta  \epsilon ^2 \pi }{M^2 t^2}}
\text{ArcTan}\bigg(\frac{m}{t \sqrt{\delta } \sqrt{\pi }}\bigg)}{m
\sqrt{\frac{m^2 \epsilon }{M^2 t^2}} \sqrt{\delta \epsilon  \pi
}}+\frac{2 \sqrt{3} \sqrt{K \delta  \epsilon  \pi }
\text{ArcTan}\bigg(\frac{\sqrt{K} M}{\sqrt{3} \sqrt{\delta  \epsilon
\pi }}\bigg)}{K M}\bigg)\\\nonumber&+&\frac{\bigg(-6+\frac{\sqrt{3}
\sqrt{K}}{\sqrt{\frac{m^2 \epsilon }{M^2 t^2}}}\bigg) C_2}{\sqrt{K}
M}\bigg(18 \bigg(m-\frac{3 m \sqrt{\delta  \epsilon  \pi } \sqrt{K
\delta \epsilon  \pi }}{\sqrt{K} \bigg(K M^2+3 \delta  \epsilon \pi
\bigg)}\bigg)\bigg)^{-1}.
\end{eqnarray}
The plot of EoS parameter is shown in Figure 7 with respect to $z$.
All the trajectories of EoS parameter represent the quintessence
phase of the universe. Figure 8 shows the graph of
$\omega_{EA}-\omega'_{EA}$ plane for same range of $z$. The
trajectories of $\omega'_{EA}$ describe the negative behavior for
all $\omega_{EA}<0$ give the freezing region of the universe. To
check the stability of the underlying model, Figure 9 shows the
unstable behavior of the model. However, for $m=2$, we get some
stable points for $z<-0.475$.\\
\underline{\textbf{Exponential form of scale factor:}}\\
Taking into account second scale factor Eq.(\ref{18}) along with
$F(K)$, we get the following energy density and pressure
\begin{eqnarray}\label{55}
\rho_{EA}&=&\frac{3 C^2 t^{-2+2 \theta } \alpha ^2 \theta ^2
\bigg(-3 \sqrt{\delta  \epsilon  \pi } \sqrt{K \delta  \epsilon  \pi
}+\sqrt{K} \bigg(K M^2+3 \delta  \epsilon  \pi \bigg)\bigg)}{8
\sqrt{K} \pi  \bigg(K M^2+3 \delta  \epsilon  \pi \bigg)}
\\\nonumber p_{EA}&=&\frac{1}{24 \pi }C^2 t^{-3+\theta } \alpha
\theta  \bigg(9 t^{1+\theta } \alpha  \theta  \bigg(-1+\frac{3
\sqrt{\delta \epsilon \pi } \sqrt{K \delta  \epsilon  \pi
}}{\sqrt{K} \bigg(K M^2+3 \delta \epsilon  \pi
\bigg)}\bigg)-\frac{1}{M^2}\epsilon (-1\\\nonumber&+&\theta )
\bigg(3 M t \bigg(M \bigg(2-\frac{3 \sqrt{\delta \epsilon \pi }
\sqrt{K \delta  \epsilon  \pi }}{\sqrt{K} \bigg(K M^2+3 \delta
\epsilon  \pi \bigg)}\bigg)-\frac{ \text{ArcTan}\bigg(\frac{\sqrt{K}
M}{\sqrt{3} \sqrt{\delta \epsilon \pi
}}\bigg)}{K}\\\nonumber&\times&\sqrt{3} \sqrt{K \delta \epsilon \pi
}+\frac{C_2}{\sqrt{K}}\bigg)(\epsilon )^{-1}+\bigg(\delta \pi
\bigg(3 t^{\theta } \alpha  \delta ^\frac{3}{2} \epsilon ^2 \theta
\pi ^\frac{3}{2} \bigg(t^{1+\theta } \alpha \sqrt{\delta } \theta
\sqrt{\pi }\\\nonumber&\times& \bigg(t^{2 \theta } \alpha ^2 \theta
^2-t^2 \delta  \pi \bigg)+\bigg(t^{2 \theta } \alpha ^2 \theta
^2+t^2 \delta  \pi \bigg)^2 \text{ArcTan}\bigg(\frac{t^{-1+\theta }
\alpha \theta }{\sqrt{\delta } \sqrt{\pi
}}\bigg)\bigg)\\\nonumber&-&\sqrt{3} M t \sqrt{\delta \epsilon  \pi
} \sqrt{\frac{t^{-2+2 \theta } \alpha ^2 \delta \epsilon ^2 \theta
^2 \pi }{M^2}} \bigg(t^{2 \theta } \alpha ^2 \theta ^2+t^2 \delta
\pi \bigg)^2
C_2\bigg)\bigg)/\bigg(\alpha\theta\\\label{56}&\times&\sqrt{\frac{t^{-2+2
\theta } \epsilon}{M^2}} (\delta  \epsilon  \pi )^\frac{3}{2}
\sqrt{\frac{t^{-2+2 \theta } \alpha ^2 \delta \epsilon ^2 \theta ^2
\pi }{M^2}} \bigg(t^{2 \theta } \alpha ^2 \theta ^2+t^2 \delta  \pi
\bigg)^2\bigg)\bigg)\bigg).
\end{eqnarray}
\begin{figure}[h]
\begin{minipage}{14pc}
\includegraphics[width=16pc]{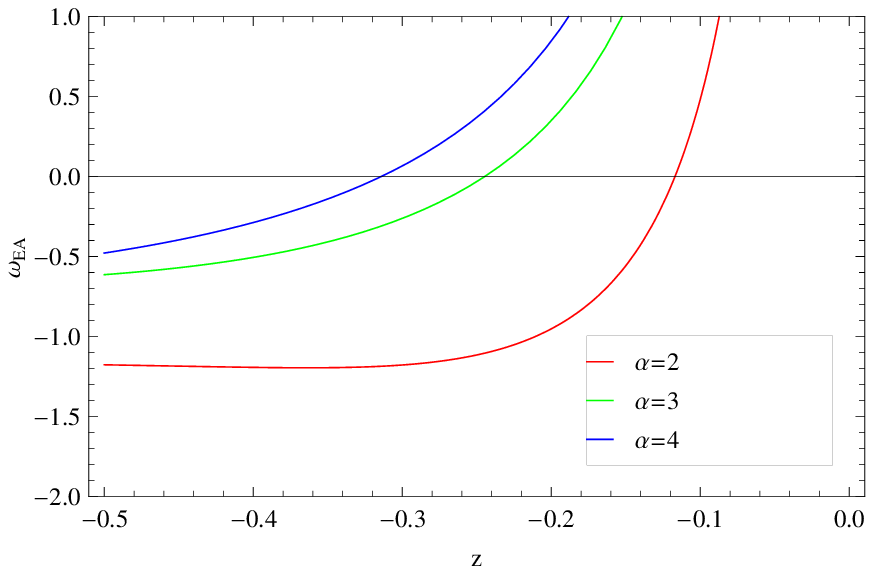}\caption{Plot of
$\omega_{EA}$ versus $z$ taking exponential scale factor for RHDE
model.}
\end{minipage}\hspace{3pc}
\begin{minipage}{14pc}
\includegraphics[width=16pc]{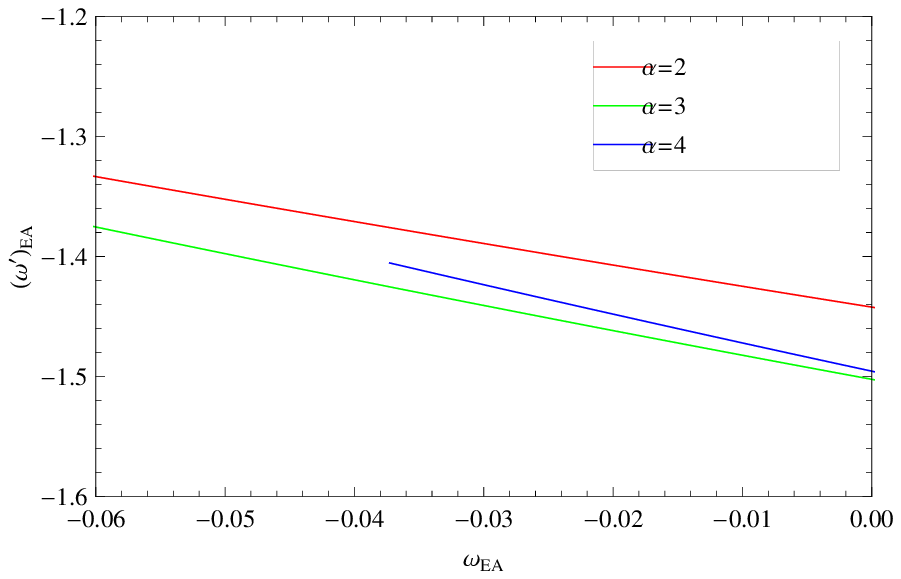}\caption{Plot of
$\omega'_{EA}-\omega_{EA}$ taking exponential scale factor for RHDE
model.}
\end{minipage}\hspace{3pc}
\begin{minipage}{14pc}
\includegraphics[width=16pc]{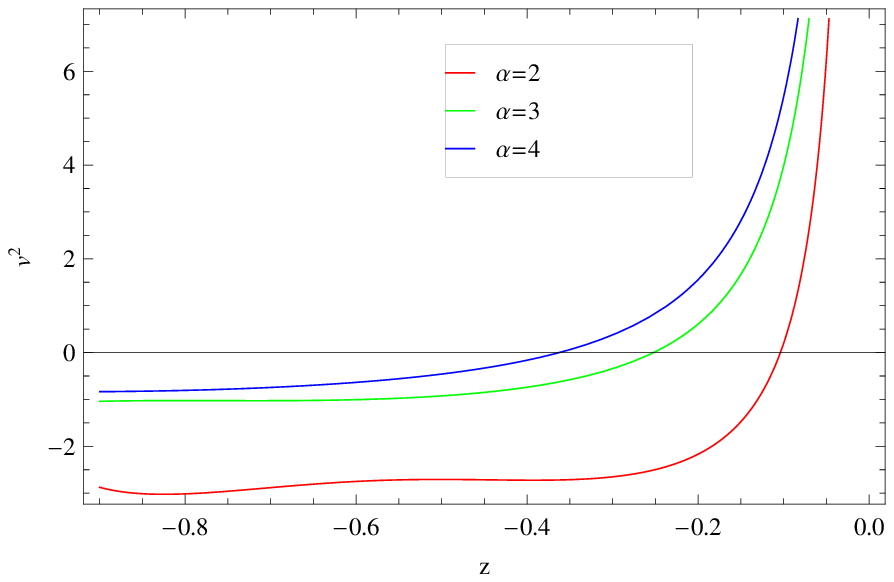}\caption{Plot of $v_s^2$
versus $z$ taking exponential scale factor for RHDE model.}
\end{minipage}\hspace{3pc}
\end{figure}

The EoS parameter is obtained from above energy density and
pressure. This parameter with its derivative are given by
\begin{eqnarray}\nonumber
\omega_{EA}&=&\bigg(\sqrt{K} t^{-1-\theta } \bigg(K M^2+3 \delta
\epsilon  \pi \bigg) \bigg(9 t^{1+\theta } \alpha  \theta
\bigg(-1+\frac{3 \sqrt{\delta  \epsilon  \pi } \sqrt{K \delta
\epsilon  \pi }}{\sqrt{K} \bigg(K M^2+3 \delta  \epsilon  \pi
\bigg)}\bigg)\\\nonumber&-&\frac{1}{M^2}\epsilon  (-1+\theta )
\bigg(3 M t \bigg(M \bigg(2-\frac{3 \sqrt{\delta  \epsilon \pi }
\sqrt{K \delta \epsilon  \pi }}{\sqrt{K} \bigg(K M^2+3 \delta
\epsilon \pi \bigg)}\bigg)-\sqrt{3} \sqrt{K \delta  \epsilon \pi
}\\\nonumber&\times&\frac{\text{ArcTan}\bigg(\frac{\sqrt{K}
M}{\sqrt{3} \sqrt{\delta \epsilon \pi }}\bigg)}{K}
+\frac{C_2}{\sqrt{K}}\bigg)(\epsilon )^{-1}+\bigg(\delta \pi \bigg(3
t^{\theta } \alpha  \delta ^\frac{3}{2} \epsilon ^2 \theta \pi
^\frac{3}{2} \bigg(t^{1+\theta } \alpha \sqrt{\delta
}\\\nonumber&\times&\theta \sqrt{\pi } \bigg(t^{2 \theta } \alpha ^2
\theta ^2-t^2 \delta  \pi \bigg)+\bigg(t^{2 \theta } \alpha ^2
\theta ^2+t^2 \delta  \pi \bigg)^2
\text{ArcTan}\bigg(\frac{t^{-1+\theta } \alpha  \theta
}{\sqrt{\delta } \sqrt{\pi }}\bigg)\bigg)\\\nonumber&-&\sqrt{3} M t
\sqrt{\delta \epsilon  \pi } \sqrt{\frac{t^{-2+2 \theta } \alpha ^2
\delta \epsilon ^2 \theta ^2 \pi }{M^2}} \bigg(t^{2 \theta } \alpha
^2 \theta ^2+t^2 \delta  \pi \bigg)^2
C_2\bigg)\bigg)/\bigg(\alpha\theta\sqrt{\epsilon}\\\nonumber&\times&\sqrt{\frac{t^{-2+2
\theta } }{M^2}} (\delta  \epsilon  \pi )^\frac{3}{2}
\sqrt{\frac{t^{-2+2 \theta } \alpha ^2 \delta  \epsilon ^2 \theta ^2
\pi }{M^2}} \bigg(t^{2 \theta } \alpha ^2 \theta ^2+t^2 \delta  \pi
\bigg)^2\bigg)\bigg)\bigg)\bigg)/\bigg(9 \\\label{57}&\times&\alpha
\theta \bigg(-3 \sqrt{\delta  \epsilon  \pi } \sqrt{K \delta
\epsilon  \pi }+\sqrt{K} \bigg(K M^2+3 \delta  \epsilon  \pi
\bigg)\bigg)\bigg).
\\\nonumber
{\omega'_{EA}}&=&\bigg(t^{-5 \theta } (-1+\theta ) \bigg(K M^2+3
\delta  \epsilon  \pi \bigg) \bigg(3 \bigg(-\frac{3 \sqrt{K} M t^{3
\theta } \alpha ^3 \theta ^4 \sqrt{\delta  \epsilon  \pi } \sqrt{K
\delta  \epsilon  \pi }}{K M^2+3 \delta  \epsilon  \pi
}\\\nonumber&+&\frac{1}{\epsilon ^2 \bigg(t^{2 \theta } \alpha ^2
\theta ^2+t^2 \delta  \pi \bigg)^3}K M^3 t^{2+\theta } \alpha \theta
\sqrt{\frac{t^{-2+2 \theta } \alpha ^2 \epsilon  \theta ^2}{M^2}}
\bigg(2 t^{6 \theta } \alpha ^6 \epsilon  \theta ^7
\alpha\theta\\\nonumber&\times&\sqrt{\frac{t^{-2+2 \theta
}\epsilon}{M^2}}+t^6 \delta ^2 \pi ^2 \bigg(2 \delta  \epsilon
\theta \sqrt{\frac{t^{-2+2 \theta } \alpha ^2 \epsilon  \theta
^2}{M^2}} \pi +(1-2 \theta ) \sqrt{\delta \epsilon  \pi }
\alpha\epsilon\theta\\\nonumber&\times&\sqrt{\frac{t^{-2+2 \theta }
\delta\pi }{M^2}}\bigg)+2 t^{4+2 \theta } \alpha ^2 \delta  \theta
^2 \pi \bigg(3 \delta \epsilon \theta \sqrt{\frac{t^{-2+2 \theta }
\alpha ^2 \epsilon \theta ^2}{M^2}} \pi -4 (-1\\\nonumber&+&\theta )
\sqrt{\delta \epsilon  \pi } \sqrt{\frac{t^{-2+2 \theta } \alpha ^2
\delta \epsilon ^2 \theta ^2 \pi }{M^2}}\bigg)+t^{2+4 \theta }
\alpha ^4 \theta ^4 \bigg(6 \delta \epsilon  \theta
\sqrt{\frac{t^{-2+2 \theta } \alpha ^2 \epsilon \theta ^2}{M^2}} \pi
\\\nonumber&+&(-1+2 \theta ) \sqrt{\delta \epsilon  \pi } \sqrt{\frac{t^{-2+2
\theta } \alpha ^2 \delta \epsilon ^2 \theta ^2 \pi
}{M^2}}\bigg)\bigg)-\delta ^{3/2} \pi ^{3/2} \bigg(\sqrt{3} t^{3
\theta } \alpha ^3 \epsilon ^{3/2} \theta ^4\\\nonumber&\times&
\sqrt{K \delta \epsilon \pi } \text{ArcTan}\bigg(\frac{\sqrt{K}
M}{\sqrt{3} \sqrt{\delta } \sqrt{\epsilon } \sqrt{\pi }}\bigg)+K M^3
t^3 (1-2 \theta ) \sqrt{\frac{t^{-2+2 \theta } \alpha ^2 \epsilon
\theta ^2}{M^2}}\\\nonumber&\times& \sqrt{\frac{t^{-2+2 \theta }
\alpha ^2 \delta \epsilon ^2 \theta ^2 \pi }{M^2}}
\text{ArcTan}\bigg(\frac{t^{-1+\theta } \alpha \theta }{\sqrt{\delta
} \sqrt{\pi }}\bigg)\bigg)((\delta  \epsilon \pi
)^{3/2})^{-1}\bigg)+\alpha  \theta  \bigg(3\\\nonumber&\times&
\sqrt{K} t^{3 \theta } \alpha ^2 \epsilon  \theta ^3-\sqrt{3} K M^2
t^{2+\theta } \sqrt{\frac{t^{-2+2 \theta } \alpha ^2 \epsilon \theta
^2}{M^2}} (-1+2 \theta )\bigg) C_2(\epsilon
)^{-1}\bigg)\\\label{58}&\times&\bigg)/\bigg(9 \sqrt{K} M \alpha ^5
\theta ^5 \bigg(-3 \sqrt{\delta  \epsilon  \pi } \sqrt{K \delta
\epsilon \pi }+\sqrt{K} \bigg(K M^2+3 \delta \epsilon  \pi
\bigg)\bigg)\bigg).
\end{eqnarray}
The correspond expression for $v^2_s$ is given by
\begin{eqnarray}\nonumber
v_s^2&=&\bigg(t^{-\theta } \bigg(K M^2+3 \delta  \epsilon  \pi
\bigg) \bigg(3 \bigg(\frac{3 \sqrt{K} M \bigg(-2+\theta +6 t^{\theta
} \alpha  \theta \bigg) \sqrt{\delta  \epsilon  \pi } \sqrt{K \delta
\epsilon  \pi }}{K M^2+3 \delta  \epsilon  \pi }\\\nonumber&+&K M
\bigg(4-\frac{t^6 \delta ^3 \pi ^3 \sqrt{\frac{t^{-2+2 \theta }
\alpha ^2 \delta  \epsilon ^2 \theta ^2 \pi
}{M^2}}}{\sqrt{\frac{t^{-2+2 \theta } \alpha ^2 \epsilon  \theta
^2}{M^2}} \sqrt{\delta  \epsilon  \pi } \bigg(t^{2 \theta } \alpha
^2 \theta ^2+t^2 \delta  \pi \bigg)^3}+t^{6 \theta } \alpha ^6
\theta ^6 \sqrt{\delta  \epsilon  \pi
}\\\nonumber&\times&\frac{\sqrt{\frac{t^{-2+2 \theta } \alpha ^2
\delta \epsilon ^2 \theta ^2 \pi }{M^2}}}{M^2 \bigg(\frac{t^{-2+2
\theta } \alpha ^2 \epsilon  \theta ^2}{M^2}\bigg)^{3/2} \bigg(t^{2
\theta } \alpha ^2 \theta ^2+t^2 \delta  \pi \bigg)^3}+8 M^2 t^6
\sqrt{\frac{t^{-2+2 \theta } \alpha ^2 \epsilon  \theta ^2}{M^2}}
\\\nonumber&\times&(\delta  \epsilon  \pi )^\frac{3}{2}\frac{ \sqrt{\frac{t^{-2+2
\theta } \alpha ^2 \delta  \epsilon ^2 \theta ^2 \pi
}{M^2}}}{\epsilon ^3 \bigg(t^{2 \theta } \alpha ^2 \theta ^2+t^2
\delta  \pi \bigg)^3}+2 \theta \bigg(-1-3 t^{\theta } \alpha -4 M^2
t^6\alpha\theta(\delta \epsilon \pi )^\frac{3}{2}\\\nonumber&\times&
\sqrt{\frac{t^{-2+2 \theta }\epsilon}{M^2}} \frac{
\sqrt{\frac{t^{-2+2 \theta } \alpha ^2 \delta  \epsilon ^2 \theta ^2
\pi }{M^2}}}{\epsilon ^3 \bigg(t^{2 \theta } \alpha ^2 \theta ^2+t^2
\delta  \pi \bigg)^3}\bigg)\bigg)+K t^{-1+\theta } \alpha \theta
\sqrt{\delta \epsilon  \pi } \alpha\epsilon\theta\sqrt{\frac{t^{-2+2
\theta } \delta \pi }{M^2}}\\\nonumber&\times&\frac{
\text{ArcTan}\bigg(\frac{t^{-1+\theta } \alpha \theta }{\sqrt{\delta
} \sqrt{\pi }}\bigg)}{M \sqrt{\delta } \bigg(\frac{t^{-2+2 \theta }
\alpha ^2 \epsilon  \theta ^2}{M^2}\bigg)^{3/2} \sqrt{\pi
}}+\sqrt{3} (-2+\theta ) \sqrt{K \delta  \epsilon  \pi }
\text{ArcTan}\bigg(\frac{\sqrt{K} M}{\sqrt{3} \sqrt{\delta  \epsilon
\pi }}\bigg)\bigg)\\\nonumber&+&\sqrt{K} \bigg(6-3 \theta
-\frac{\sqrt{3} \sqrt{K}}{\sqrt{\frac{t^{-2+2 \theta } \alpha ^2
\epsilon  \theta ^2}{M^2}}}\bigg) C_2\bigg)\bigg)/\bigg(18 \sqrt{K}
M \alpha  \theta \bigg(-3 \sqrt{\delta  \epsilon  \pi } \sqrt{K
\delta  \epsilon  \pi }\\\label{60}&+&\sqrt{K} \bigg(K M^2+3 \delta
\epsilon \pi \bigg)\bigg)\bigg).
\end{eqnarray}

Figure 10 represents the graph of EoS parameter versus $z$ for RHDE
model taking exponential form of the scale factor. For $\alpha=2$,
initially the trajectory expresses the transition from decelerated
phase to accelerated phase and then crosses the phantom divide line
and gives the phantom phase of the universe. For higher values of
the $\alpha$ that is for $\alpha=3,4$, the trajectories of EoS
parameter represents the quintessence phase. In Figure 11, we plot
the graph of evolution parameter of EoS versus EoS parameter which
gives the freezing region of the universe. Figure 12 shows the graph
of $v_s^{2}$ for stability analysis of the model. Initially the
graph gives the stability and then for decreasing $z$, the model
becomes unstable. As we increase the value of $\alpha$, the
trajectories give more stable points.

\section{Reconstruction from Sharma-Mittal Holographic Dark Eneryg Model}

Sharma-Mittal introduced a two parametric entropy and is defined as
\cite{30ss}
\begin{equation}\label{17}
S_{SM}=\frac{1}{1-r}\left((\Sigma{_{i=1}^{n}}
P{_{i}^{1-\delta}})^{1-r/\delta}-1\right),
\end{equation}
where $r$ is a new free parameter. The expression of SMHDE model for
Hubble horizon is given by
\begin{equation}\label{521}
\rho_{D}=\frac{3\epsilon H^4}{8\pi R}\bigg((1+\frac{\delta
\pi}{H^2})^\frac{R}{\delta}-1\bigg).
\end{equation}
By comparing the energy densities of SMHDE model and Einstein-Aether
gravity model, we find
\begin{equation}\label{532}
\frac{dF}{dK}-\frac{F}{2K}=\frac{KM^2}{24\epsilon\pi
R}\bigg(\bigg(1+\frac{3\epsilon\pi\delta}{KM^2}\bigg)^\frac{R}{\delta}-1\bigg),
\end{equation}
which leads us to the following solution
\begin{equation}\label{54}
F(K)=\frac{K^2M^2\bigg(-1+{_2}F_1(-\frac{3}{2},-\frac{R}{\delta},\frac{-1}{2},\frac{-3\pi
\delta \epsilon}{KM^2}) \bigg)}{36\pi R \epsilon}+C_3\sqrt{K}.
\end{equation}
\underline{\textbf{Power-law form of scale factor:}}\\
For this scale factor, we obtain
\begin{eqnarray}\label{62}\rho_{EA}&=&\frac{K m^2 M^2 \bigg(-1+\bigg(1+\frac{3 \pi  \delta
\epsilon }{K M^2}\bigg)^\frac{R}{\delta}\bigg)}{8 \pi  R t^2}.
\\\nonumber p_{EA}&=&\frac{1}{72 t^4}m \bigg(\frac{1}{\pi  R
\bigg(m^2+\pi  t^2 \delta \bigg)}\bigg(3 m^2 \bigg(-8 \bigg(m^2+\pi
t^2 \delta \bigg)\\\nonumber&+&3 \bigg(1+\frac{\pi  t^2 \delta
}{m^2}\bigg)^\frac{R}{\delta} \bigg(3 m^2+\pi t^2 (-2 R+3 \delta
)\bigg)\bigg) \epsilon +K M^2\\\nonumber&\times& t^2 \bigg(m^2+\pi
t^2 \delta \bigg) \bigg(-4+9 m-3 (-1+3 m) \bigg(1+\frac{3 \pi \delta
\epsilon }{K
M^2}\bigg)^\frac{R}{\delta}\bigg)\\\nonumber&-&\bigg(m^2+\pi t^2
\delta \bigg) \bigg(3 m^2 \epsilon
{_2}F_1\bigg(-\frac{3}{2},-\frac{R}{\delta },-\frac{1}{2},-\frac{\pi
t^2 \delta }{m^2}\bigg)\\\nonumber&-&K M^2 t^2
{_2}F_1\bigg(-\frac{3}{2},-\frac{R}{\delta },-\frac{1}{2},-\frac{3
\pi  \delta  \epsilon }{K
M^2}\bigg)\bigg)\bigg)\\\label{63}&-&\frac{12 t^2 \epsilon
\bigg(-3+\frac{\sqrt{3} \sqrt{K}}{\sqrt{\frac{m^2 \epsilon }{M^2
t^2}}}\bigg) C_3}{\sqrt{K}}\bigg).
\end{eqnarray}
The cosmological parameters are given by
\begin{eqnarray}\nonumber
\omega_{EA}&=&\frac{1}{9 K m M^2 t^2 \bigg(-1+\bigg(1+\frac{3 \pi
\delta  \epsilon }{K M^2}\bigg)^\frac{R}{\delta}\bigg)}\pi  R
\bigg(\frac{1}{\pi  R \bigg(m^2+\pi  t^2 \delta \bigg)}\bigg(3
m^2\\\nonumber&\times& \bigg(-8 \bigg(m^2+\pi t^2 \delta \bigg)+3
\bigg(1+\frac{\pi  t^2 \delta }{m^2}\bigg)^\frac{R}{\delta} \bigg(3
m^2+\pi  t^2 (-2 R+3 \delta )\bigg)\bigg)\\\nonumber&\times&
\epsilon +K M^2 t^2 \bigg(m^2+\pi  t^2 \delta \bigg) \bigg(-4+9 m-3
(-1+3 m) \bigg(1+\frac{3 \pi  \delta \epsilon }{K
M^2}\bigg)^\frac{R}{\delta}\\\nonumber&\times&\bigg)-\bigg(m^2+\pi
t^2 \delta \bigg) \bigg(3 m^2 \epsilon
{_2}F_1\bigg(-\frac{3}{2},-\frac{R}{\delta },-\frac{1}{2},-\frac{\pi
t^2 \delta }{m^2}\bigg)\\\nonumber&-&K M^2 t^2
{_2}F_1\bigg(-\frac{3}{2},-\frac{R}{\delta },-\frac{1}{2},-\frac{3
\pi  \delta  \epsilon }{K M^2}\bigg)\bigg)\bigg)-12 t^2
\epsilon\\\label{67}&\times&\frac{ \bigg(-3+\frac{\sqrt{3}
\sqrt{K}}{\sqrt{\frac{m^2 \epsilon }{M^2 t^2}}}\bigg)
C_3}{\sqrt{K}}\bigg), \\\nonumber {\omega'_{EA}}&=&m^4 \epsilon
\bigg(-3 \bigg(1+\frac{\pi  t^2 \delta }{m^2}\bigg)^\frac{R}{\delta}
\bigg(5 m^4+2 m^2 \pi  t^2 (-3 R+5 \delta )+\pi ^2 t^4 \bigg(4
R^2\\\nonumber&-&10 R \delta +5 \delta ^2\bigg)\bigg)-\bigg(m^2+\pi
t^2 \delta \bigg)^2
\bigg(-16\\\nonumber&+&{_2}F_1\bigg(-\frac{3}{2},-\frac{R}{\delta
},-\frac{1}{2},-\frac{\pi  t^2 \delta }{m^2}\bigg)\bigg)\bigg)-4
\sqrt{3} M^2 \pi  R t^4 \\\nonumber&\times&\bigg(m^2+\pi  t^2 \delta
\bigg)^2 \sqrt{\frac{m^2 \epsilon }{M^2 t^2}} C_3\bigg(3 K m^4 M^2
t^2 \bigg(m^2+\pi  t^2 \delta \bigg)^2
\bigg(-1+\bigg(1\\\label{69}&+&\frac{3 \pi \delta  \epsilon }{K
M^2}\bigg)^\frac{R}{\delta}\bigg)\bigg), \\\nonumber
v_s^2&=&\frac{1}{18 K^\frac{3}{2} m^3 M^2 t^2 \bigg(m^2+\pi  t^2
\delta \bigg)^2 \bigg(-1+\bigg(1+\frac{3 \pi  \delta  \epsilon }{K
M^2}\bigg)^{R/\delta }\bigg)}\bigg(\sqrt{K} m^2
\\\nonumber&\times&\bigg(3 m^2 \bigg(-32 \bigg(m^2+\pi  t^2 \delta
\bigg)^2+3 \bigg(1+\frac{\pi t^2 \delta }{m^2}\bigg)^{R/\delta }
\bigg(11 m^4+2 m^2 \pi  t^2\\\nonumber&\times& (-5 R+11 \delta )+\pi
^2 t^4 \bigg(4 R^2-14 R \delta +11 \delta ^2\bigg)\bigg)\bigg)
\epsilon -2 K M^2 t^2 \bigg(m^2\\\nonumber&+&\pi  t^2 \delta
\bigg)^2 \bigg(4-9 m+3 (-1+3 m) \bigg(1+\frac{3 \pi  \delta \epsilon
}{K M^2}\bigg)^{R/\delta }\bigg)-\bigg(m^2+\pi  t^2 \delta \bigg)^2
\\\nonumber&\times&\bigg(3 m^2 \epsilon
{_2}F_1\bigg(-\frac{3}{2},-\frac{R}{\delta },-\frac{1}{2},-\frac{\pi
t^2 \delta }{m^2}\bigg)-2 K M^2 t^2\\\nonumber&\times&
{_2}F_1\bigg(-\frac{3}{2},-\frac{R}{\delta },-\frac{1}{2},-\frac{3
\pi  \delta  \epsilon }{K M^2}\bigg)\bigg)\bigg)+12 \pi  R t^2
\bigg(m^3\\\label{70}&+&m \pi t^2 \delta \bigg)^2 \epsilon
\bigg(6-\frac{\sqrt{3} \sqrt{K}}{\sqrt{\frac{m^2 \epsilon }{M^2
t^2}}}\bigg) C_3\bigg),
\end{eqnarray}
\begin{figure}[h]
\begin{minipage}{14pc}
\includegraphics[width=16pc]{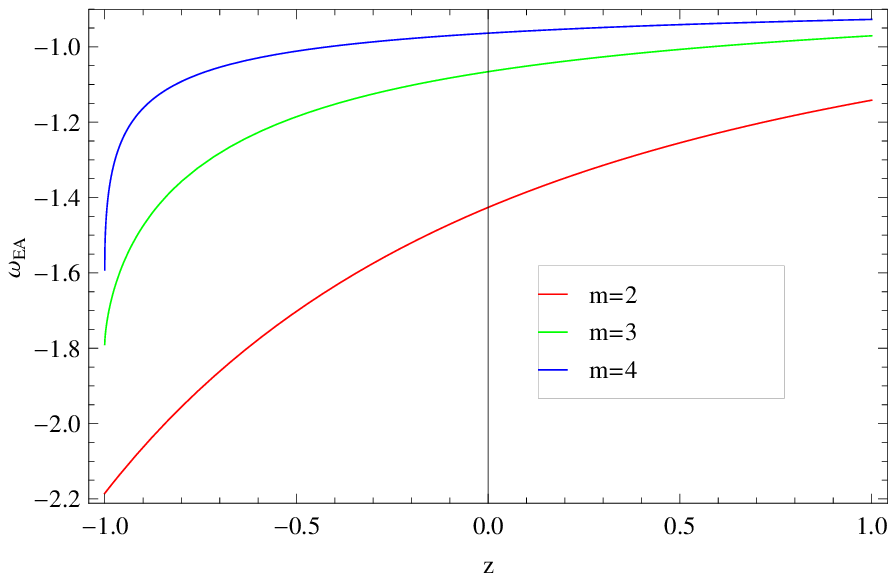}\caption{Plot of
$\omega_{EA}$ versus $z$ taking power-law scale factor for SMHDE
model.}
\end{minipage}\hspace{3pc}
\begin{minipage}{14pc}
\includegraphics[width=16pc]{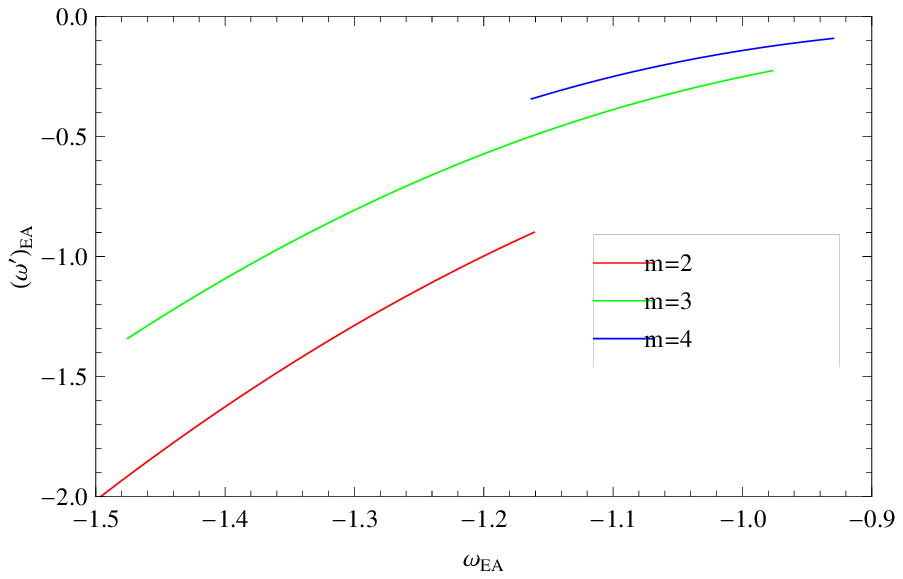}\caption{Plot of
$\omega'_{EA}-\omega_{EA}$ taking power-law scale factor for SMHDE
model.}
\end{minipage}\hspace{3pc}
\begin{minipage}{14pc}
\includegraphics[width=16pc]{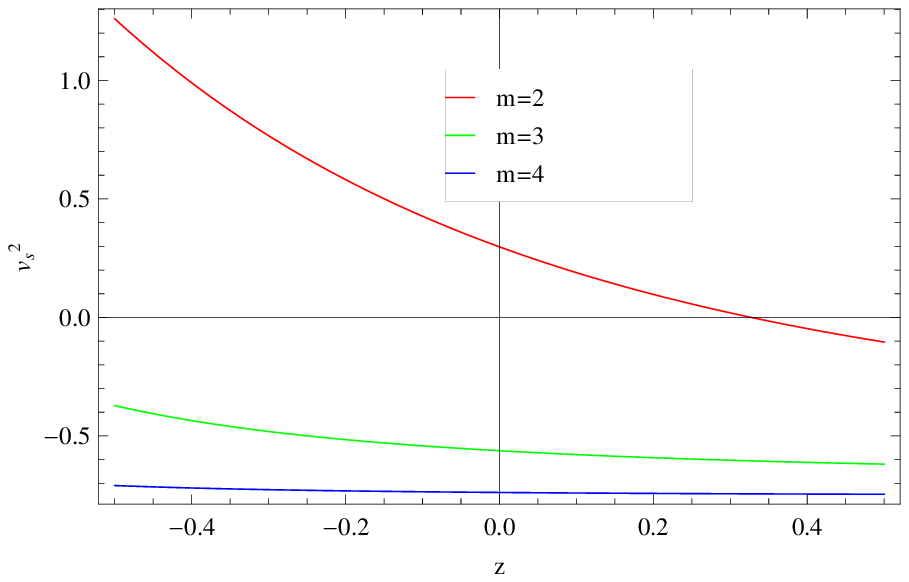}\caption{Plot of $v_s^2$
versus $z$ taking power-law scale factor for SMHDE model.}
\end{minipage}\hspace{3pc}
\end{figure}

We plot EoS parameter for SMHDE model with respect to redshift
parameter as shown in Figure 13 for power-law scale factor. For
$m=3$ and $4$, the trajectories represent the transition from
quintessence to phantom phase while $m=2$ indicates the phantom era
throughout for $z$. The plot of this parameter with its evolution
parameter is given in Figure 14 which shows the freezing region of
the evolving universe. However, for higher values of $m$, we may get
thawing region $(\omega'_{EA}>0)$. Figure 15 gives the graph of
squared speed of sound versus redshift. The trajectory for $m=2$
shows the stability of the model as redshift parameter decreases
while other trajectories describe the unstable behavior of the
model. \\
\underline{\textbf{Exponential form of scale factor:}}\\
Following the same steps, we obtain the following expressions for
energy density, pressure and parameters for exponential scale
factor. These are
\begin{eqnarray}\label{81}
\rho_{EA}&=&\frac{K M^2 t^{-2+2 \theta } \alpha ^2
\bigg(-1+\bigg(1+\frac{3 \pi  \delta  \epsilon }{K
M^2}\bigg)^{R/\delta }\bigg) \theta ^2}{8 \pi  R}, \\\nonumber
p_{EA}&=&\frac{1}{24} t^{-4+\theta } \alpha  \theta  \bigg(-\frac{3
K M^2 t^{2+\theta } \alpha  \bigg(-1+\bigg(1+\frac{3 \pi  \delta
\epsilon }{K M^2}\bigg)^{R/\delta }\bigg) \theta }{\pi R}-\epsilon
(-1+\theta )\\\nonumber&\times& \bigg(t^2 \bigg(K^{3/2} M^2
\bigg(-4+3 \bigg(1+\frac{3 \pi  \delta  \epsilon }{K
M^2}\bigg)^{R/\delta
}\\\nonumber&+&{_2}F_1\bigg(-\frac{3}{2},-\frac{R}{\delta
},-\frac{1}{2},-\frac{3 \pi  \delta  \epsilon }{K
M^2}\bigg)\bigg)+36 \pi  R \epsilon  C_3\bigg)\\\nonumber&\times&(3
\sqrt{K} \pi  R \epsilon )^{-1}+t^{2 \theta } \alpha ^2 \theta ^2
\bigg(-8-3 \bigg(1+\frac{\pi t^{2-2 \theta } \delta }{\alpha ^2
\theta ^2}\bigg)^{R/\delta } \bigg(\pi  t^2 (-2 R\\\nonumber&+&3
\delta )+3 t^{2 \theta } \alpha ^2 \theta ^2\bigg)(\pi  t^2 \delta
+t^{2 \theta } \alpha ^2 \theta
^2)^{-1}\\\nonumber&+&{_2}F_1\bigg(-\frac{3}{2},-\frac{R}{\delta
},-\frac{1}{2},-\frac{\pi  t^{2-2 \theta } \delta }{\alpha ^2 \theta
^2}\bigg)(\pi  R)^{-1}-4 \sqrt{3} \epsilon\\\label{82}&\times&\frac{
C_3}{M^2 \bigg(\frac{t^{-2+2 \theta } \alpha ^2 \epsilon  \theta
^2}{M^2}\bigg)^{3/2}}\bigg)\bigg)\bigg), \\\nonumber
\omega_{EA}&=&\frac{1}{3 K M^2 \alpha  \bigg(-1+\bigg(1+\frac{3 \pi
\delta  \epsilon }{K M^2}\bigg)^{R/\delta }\bigg) \theta }\pi  R
t^{-2-\theta } \bigg(-3 K M^2 t^{2+\theta } \alpha
\bigg(-1\\\nonumber&+&\bigg(1+\frac{3 \pi  \delta  \epsilon }{K
M^2}\bigg)^{R/\delta }\bigg) \theta (\pi  R)^{-1}-\epsilon
(-1+\theta ) \bigg(t^2 \bigg(K^{3/2} M^2 \bigg(-4\\\nonumber&+&3
\bigg(1+\frac{3 \pi \delta  \epsilon }{K M^2}\bigg)^{R/\delta
}+{_2}F_1\bigg(-\frac{3}{2},-\frac{R}{\delta },-\frac{1}{2},-\frac{3
\pi  \delta  \epsilon }{K M^2}\bigg)\bigg)\\\nonumber&+&36 \pi  R
\epsilon  C_3\bigg)(3 \sqrt{K} \pi R \epsilon )^{-1}+t^{2 \theta }
\alpha ^2 \theta ^2 \bigg(-8-3 \bigg(1+\frac{\pi t^{2-2 \theta }
\delta }{\alpha ^2 \theta ^2}\bigg)^{R/\delta } \bigg(\pi
t^2\\\nonumber&\times& (-2 R+3 \delta )+3 t^{2 \theta } \alpha ^2
\theta ^2\bigg)(\pi  t^2 \delta +t^{2 \theta } \alpha ^2 \theta
^2)^{-1}\\\nonumber&+&{_2}F_1\bigg(-\frac{3}{2},-\frac{R}{\delta
},-\frac{1}{2},-\frac{\pi  t^{2-2 \theta } \delta }{\alpha ^2 \theta
^2}\bigg)(\pi  R)^{-1}-4 \sqrt{3} \epsilon\\\label{85}&\times&\frac{
C_3}{M^2 \bigg(\frac{t^{-2+2 \theta } \alpha ^2 \epsilon  \theta
^2}{M^2}\bigg)^{3/2}}\bigg)\bigg)\bigg), \\\nonumber
{\omega'_{EA}}&=&\frac{1}{9 K M^2 \alpha ^2 \bigg(-1+\bigg(1+\frac{3
\pi \delta  \epsilon }{K M^2}\bigg)^{R/\delta }\bigg) \theta ^2}\pi
R t^{-2 (1+\theta )} (-1+\theta )\\\nonumber&\times&
\bigg(\frac{1}{\pi  R \bigg(\pi t^2 \delta +t^{2 \theta } \alpha ^2
\theta ^2\bigg)^2}\theta \bigg(K M^2 t^2 \bigg(-4+3 \bigg(1+\frac{3
\pi \delta  \epsilon }{K M^2}\bigg)^{R/\delta
}\bigg)\\\nonumber&\times& \bigg(\pi  t^2 \delta +t^{2 \theta }
\alpha ^2 \theta ^2\bigg)^2+3 t^{2 \theta } \epsilon  \theta \bigg(8
(-2+\theta ) \bigg(\pi  t^2 \alpha  \delta +t^{2 \theta } \alpha ^3
\theta ^2\bigg)^2-3 \alpha ^2\\\nonumber&\times& \bigg(1+\frac{\pi
t^{2-2 \theta } \delta }{\alpha ^2 \theta ^2}\bigg)^{R/\delta }
\bigg(t^{4 \theta } \alpha ^4 \theta ^4 (-5+2 \theta )+2 \pi  t^{2+2
\theta } \alpha ^2 \theta ^2 (R (3-2 \theta )\\\nonumber&+&\delta
(-5+2 \theta ))+\pi ^2 t^4 \bigg(2 R \delta  (5-4 \theta )+4 R^2
(-1+\theta )+\delta ^2 (-5+2 \theta
)\bigg)\bigg)\bigg)\\\nonumber&+&\bigg(\pi t^2 \delta +t^{2 \theta }
\alpha ^2 \theta ^2\bigg)^2 \bigg(K M^2 t^2
{_2}F_1\bigg(-\frac{3}{2},-\frac{R}{\delta },-\frac{1}{2},-\frac{3
\pi  \delta  \epsilon }{K M^2}\bigg)\\\nonumber&-&3 t^{2 \theta }
\alpha ^2 \epsilon  \theta (-1+2 \theta )
_2F_1\bigg(-\frac{3}{2},-\frac{R}{\delta },-\frac{1}{2},-\frac{\pi
t^{2-2 \theta } \delta }{\alpha ^2 \theta
^2}\bigg)\bigg)\bigg)\\\label{89}&+&12 t^2 \epsilon  \bigg(\frac{3
\theta }{\sqrt{K}}+\frac{\sqrt{3} (1-2 \theta )}{\sqrt{\frac{t^{-2+2
\theta } \alpha ^2 \epsilon  \theta ^2}{M^2}}}\bigg) C_3\bigg),
\\\nonumber v_s^2&=&\frac{1}{18 K M^2 \alpha  \bigg(-1+\bigg(1+\frac{3
\pi \delta \epsilon }{K M^2}\bigg)^{R/\delta }\bigg) \theta }\pi  R
t^{-2-\theta } \bigg(\frac{1}{\pi  R \bigg(\pi  t^2 \delta +t^{2
\theta } \alpha ^2 \theta ^2\bigg)^2}\\\nonumber&\times&\bigg(K M^2
t^2 \bigg(\pi  t^2 \delta +t^{2 \theta } \alpha ^2 \theta ^2\bigg)^2
\bigg(-8+4 \theta +18 t^{\theta } \alpha  \theta -3 \bigg(1+\frac{3
\pi  \delta \epsilon }{K M^2}\bigg)^{R/\delta }\\\nonumber&\times&
\bigg(-2+\theta +6 t^{\theta } \alpha  \theta \bigg)\bigg)+3 t^{2
\theta } \epsilon \bigg(8 (-4+3 \theta ) \bigg(\pi  t^2 \alpha
\delta  \theta +t^{2 \theta } \alpha ^3 \theta ^3\bigg)^2-3 \alpha
^2 \\\nonumber&\times&\bigg(1+\frac{\pi t^{2-2 \theta } \delta
}{\alpha ^2 \theta ^2}\bigg)^{R/\delta } \theta ^2 \bigg(t^{4 \theta
} \alpha ^4 \theta ^4 (-11+8 \theta )+2 \pi t^{2+2 \theta } \alpha
^2 \theta ^2 (R (5-4 \theta )\\\nonumber&+&\delta (-11+8 \theta
))+\pi ^2 t^4 \bigg(2 R \delta  (7-6 \theta )+4 R^2 (-1+\theta
)+\delta ^2 (-11+8 \theta )\bigg)\bigg)\bigg)\\\nonumber&-&\bigg(\pi
t^2 \delta +t^{2 \theta } \alpha ^2 \theta ^2\bigg)^2 \bigg(K M^2
t^2 (-2+\theta ) {_2}F_1\bigg(-\frac{3}{2},-\frac{R}{\delta
},-\frac{1}{2},-\frac{3 \pi  \delta  \epsilon }{K
M^2}\bigg)\\\nonumber&+&3 t^{2 \theta } \alpha ^2 \epsilon \theta ^2
{_2}F_1\bigg(-\frac{3}{2},-\frac{R}{\delta },-\frac{1}{2},-\frac{\pi
t^{2-2 \theta } \delta }{\alpha ^2 \theta ^2}\bigg)\bigg)\bigg)+12
t^2 \epsilon C_3\\\label{90}&\times& \frac{ \bigg(6-3 \theta
-\frac{\sqrt{3} \sqrt{K}}{\sqrt{\frac{t^{-2+2 \theta } \alpha ^2
\epsilon  \theta ^2}{M^2}}}\bigg) }{\sqrt{K}}\bigg).
\end{eqnarray}
\begin{figure}[h]
\begin{minipage}{14pc}
\includegraphics[width=16pc]{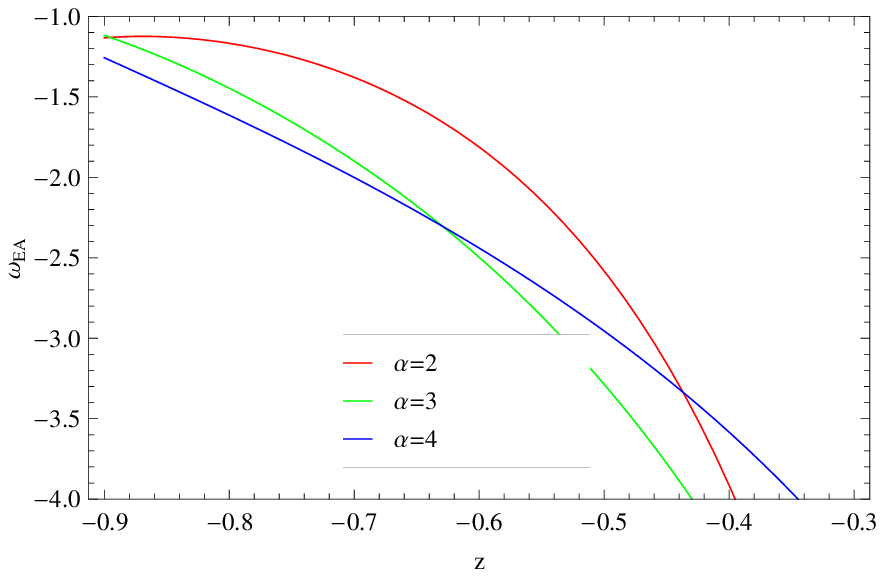}\caption{Plot of
$\omega_{EA}$ versus $z$ taking exponential scale factor for SMHDE
model.}
\end{minipage}\hspace{3pc}
\begin{minipage}{14pc}
\includegraphics[width=16pc]{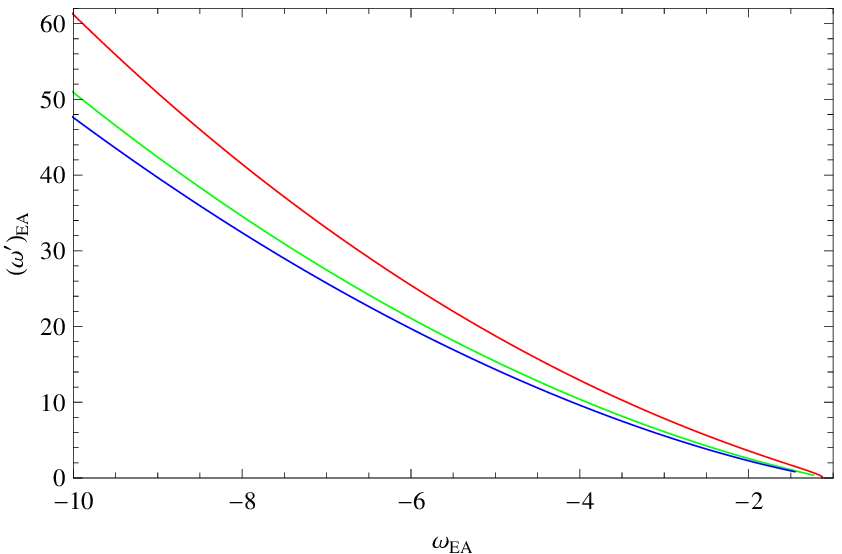}\caption{Plot of
$\omega'_{EA}-\omega_{EA}$ taking exponential scale factor for SMHDE
model.}
\end{minipage}\hspace{3pc}
\begin{minipage}{14pc}
\includegraphics[width=16pc]{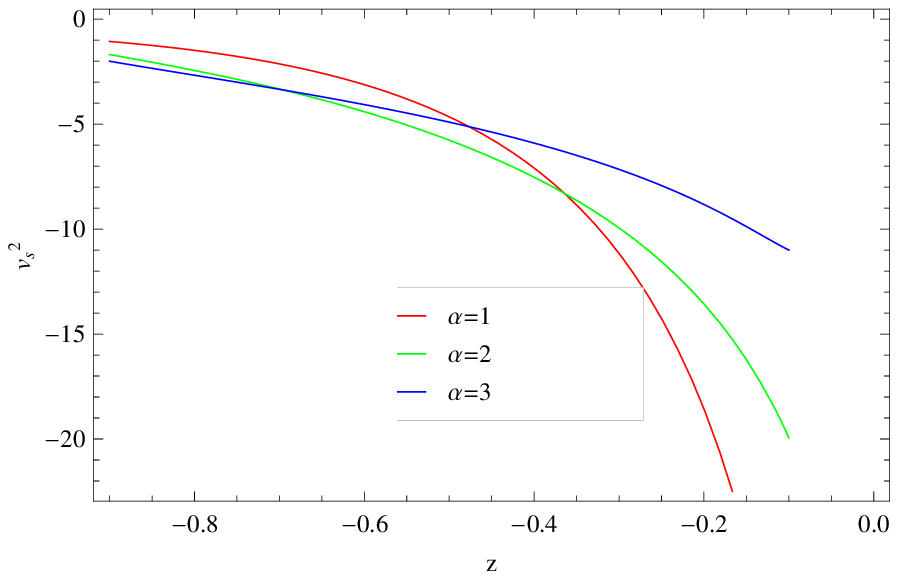}\caption{Plot of $v_s^2$
versus $z$ taking exponential scale factor for SMHDE model.}
\end{minipage}\hspace{3pc}
\end{figure}
For exponential scale factor for SMHDE model, the plot of EoS
parameter in Figure 16 represents the phantom behavior initially but
converges to cosmological constant behavior for $\alpha=2,3$ as $z$
decreases. For $\alpha=4$, the EoS parameter gives the phantom
behavior. Figure 17 represents the graph of
$\omega'_{EA}-\omega_{EA}$ plane which shows the positive behavior
of $\omega'_{EA}$ versus negative $\omega_{EA}$ expressing thawing
region of the universe. The squared speed of sound graph gives
unstable behavior of the SMHDE model in the framework of
Einstein-Aether theory of gravity.

\def\baselinestretch{1}
\section{Summary}
\def\baselinestretch{1.66}

In this work, we have discussed about the Einstein-Aether gravity
and utilized its effective density and pressure. We have developed
the Einstein-Aether models by using some holographic dark energy
models. In the presence of free function $F(K)$, we have treated the
affective density and pressure as DE. From the modified HDE models
such as THDE, RHDE and SMHDE models, we have formed the unknown
function $F(K)$ for Einstein-Aether theory by considering the
power-law form and exponential forms of scale factor. We have
discussed some cosmological parameters like, EoS parameter with its
evolutionary parameter and squared speed of sound to check the
stability of reconstructed models for this theory.

The remaining results have been summarized as follows:\\
\textbf{\underline{EoS parameter for power-law scale factor:}}
\begin{itemize}
  \item THDE $\Rightarrow$ phantom behavior,
  \item RHDE $\Rightarrow$ quintessence phase,
  \item SMHDE $\Rightarrow$ transition from
quintessence to phantom phase for $m=3, 4$, phantom era for $m=2$.
\end{itemize}
\textbf{\underline{EoS parameter for exponential scale factor:}}
\begin{itemize}
  \item THDE $\Rightarrow$ transition from quintessence to
phantom era for $\alpha=2$, phantom behavior for $\alpha=3$,
$\Lambda$CDM model for $\alpha=4$,
  \item RHDE $\Rightarrow$ phantom phase for $\alpha=2$, quintessence phase for $\alpha=3, 4$
  \item SMHDE $\Rightarrow$ cosmological constant behavior for $\alpha=2,3$, phantom behavior for $m=4$.
\end{itemize}
\textbf{\underline{$\omega'-\omega$ plane for power-law scale
factor:}}
\begin{itemize}
  \item THDE $\Rightarrow$ freezing
region for $m=2,3$, thawing region for $m=4$,
  \item RHDE $\Rightarrow$ freezing region,
  \item SMHDE $\Rightarrow$ freezing region.
\end{itemize}
\textbf{\underline{$\omega'-\omega$ plane for exponential scale
factor:}}
\begin{itemize}
  \item THDE $\Rightarrow$ freezing region to thawing region,
  \item RHDE $\Rightarrow$ freezing region,
  \item SMHDE $\Rightarrow$ thawing region.
\end{itemize}
\textbf{\underline{Squared speed of sound for power-law scale
factor:}}
\begin{itemize}
  \item THDE $\Rightarrow$ unstable,
  \item RHDE $\Rightarrow$ unstable,
  \item SMHDE $\Rightarrow$ stable for $m=2$, unstable for $m=3, 4$.
\end{itemize}
\textbf{\underline{Squared speed of sound for exponential scale
factor:}}
\begin{itemize}
  \item THDE $\Rightarrow$ unstable,
  \item RHDE $\Rightarrow$ stability for higher values and instability for lower values,
  \item SMHDE $\Rightarrow$ unstable.
\end{itemize}

It is mentioned here that for $m=2$ for power-law form of scale
factor in the case of SMHDE model, we obtain phantom region with
stable behavior in freezing region which leads to most favorable
result with current cosmic expansion scenario.\\\\

{\bf Acknowledgment}

\vspace{0.5cm}

SR and AJ is thankful to the Higher Education Commission, Islamabad,
Pakistan for its financial support under the grant No:
5412/Federal/NRPU/R\&D/HEC/2016 of NATIONAL RESEARCH PROGRAMME FOR
UNIVERSITIES (NRPU). The work of KB was partially supported by the
JSPS KAKENHI Grant Number JP 25800136 and Competitive Research Funds
for Fukushima University Faculty (18RI009).

\end{document}